%% file: ANA-STDM-2018-06-PAPER.tex
\pdfoutput=1
\newcommand*{\ATLASLATEXPATH}{}
\documentclass[PAPER, cernpreprint, texlive=2016,UKenglish,coverpage=false, LANGEDIT=true, LANGSHOW=false]{\ATLASLATEXPATH atlasdoc}
 
\usepackage{\ATLASLATEXPATH atlaspackage}
\usepackage{\ATLASLATEXPATH atlasbiblatex}
 
\usepackage{\ATLASLATEXPATH atlasphysics}

\usepackage{longtable}

\newcommand{\FidXsecCentralZmm}{ 205.6}
\newcommand{\FidXsecStatAbsZmm}{   8.1}
\newcommand{\FidXsecLumiAbsZmm}{   6.4}
\newcommand{\TotXsecCentralZmm}{ 326.3}
\newcommand{\TotXsecStatAbsZmm}{  12.9}
\newcommand{\TotXsecExtrapUncertAbsZmm}{   5.5}
\newcommand{\TotXsecLumiAbsZmm}{  10.1}
\newcommand{\FidXsecSystAbsZmm}{   2.6}
\newcommand{\TotXsecSystAbsZmm}{   4.1}
\newcommand{\FidXsecCentralWmnminus}{ 799}
\newcommand{\FidXsecStatAbsWmnminus}{  17}
\newcommand{\FidXsecLumiAbsWmnminus}{  25}
\newcommand{\TotXsecCentralWmnminus}{1402}
\newcommand{\TotXsecStatAbsWmnminus}{  30}
\newcommand{\TotXsecExtrapUncertAbsWmnminus}{  21}
\newcommand{\TotXsecLumiAbsWmnminus}{  44}
\newcommand{\FidXsecSystAbsWmnminus}{  11}
\newcommand{\TotXsecSystAbsWmnminus}{  19}
\newcommand{\FidXsecCentralWmnplus}{1438}
\newcommand{\FidXsecStatAbsWmnplus}{  23}
\newcommand{\FidXsecLumiAbsWmnplus}{  45}
\newcommand{\TotXsecCentralWmnplus}{2319}
\newcommand{\TotXsecStatAbsWmnplus}{  36}
\newcommand{\TotXsecExtrapUncertAbsWmnplus}{  30}
\newcommand{\TotXsecLumiAbsWmnplus}{  72}
\newcommand{\FidXsecSystAbsWmnplus}{  19}
\newcommand{\TotXsecSystAbsWmnplus}{  30}
\newcommand{\FidXsecCentralZee}{ 197.6}
\newcommand{\FidXsecStatAbsZee}{   9.6}
\newcommand{\FidXsecLumiAbsZee}{   6.1}
\newcommand{\TotXsecCentralZee}{ 313.6}
\newcommand{\TotXsecStatAbsZee}{  15.2}
\newcommand{\TotXsecExtrapUncertAbsZee}{   5.3}
\newcommand{\TotXsecLumiAbsZee}{   9.7}
\newcommand{\FidXsecSystAbsZee}{   9.5}
\newcommand{\TotXsecSystAbsZee}{   15.0}
\newcommand{\FidXsecCentralWenminus}{ 789}
\newcommand{\FidXsecStatAbsWenminus}{  18}
\newcommand{\FidXsecLumiAbsWenminus}{  25}
\newcommand{\TotXsecCentralWenminus}{1385}
\newcommand{\TotXsecStatAbsWenminus}{  31}
\newcommand{\TotXsecExtrapUncertAbsWenminus}{  21}
\newcommand{\TotXsecLumiAbsWenminus}{  43}
\newcommand{\FidXsecSystAbsWenminus}{   20}
\newcommand{\TotXsecSystAbsWenminus}{  36}
\newcommand{\FidXsecCentralWenplus}{1416}
\newcommand{\FidXsecStatAbsWenplus}{  24}
\newcommand{\FidXsecLumiAbsWenplus}{  44}
\newcommand{\TotXsecCentralWenplus}{2284}
\newcommand{\TotXsecStatAbsWenplus}{  38}
\newcommand{\TotXsecExtrapUncertAbsWenplus}{  30}
\newcommand{\TotXsecLumiAbsWenplus}{  71}
\newcommand{\FidXsecSystAbsWenplus}{  36}
\newcommand{\TotXsecSystAbsWenplus}{  58}
\newcommand{\AveCentralZ}{ 203.7}
\newcommand{\AveStatZ}{   6.2}
\newcommand{\AveSystZ}{   3.2}
\newcommand{\AveLumiZ}{   6.3}
\newcommand{\AveTotCentralZ}{ 323.4}
\newcommand{\AveTotStatZ}{   9.8}
\newcommand{\AveTotSystZ}{   5.0}
\newcommand{\AveTotExtrapZ}{   5.5}
\newcommand{\AveTotLumiZ}{  10.0}
\newcommand{\AveCentralWm}{ 798}
\newcommand{\AveStatWm}{  12}
\newcommand{\AveSystWm}{10}
\newcommand{\AveLumiWm}{  25}
\newcommand{\AveTotCentralWm}{1399}
\newcommand{\AveTotStatWm}{  21}
\newcommand{\AveTotSystWm}{  17}
\newcommand{\AveTotExtrapWm}{  21}
\newcommand{\AveTotLumiWm}{  43}
\newcommand{\AveCentralWp}{1433}
\newcommand{\AveStatWp}{  16}
\newcommand{\AveSystWp}{  17}
\newcommand{\AveLumiWp}{  44}
\newcommand{\AveTotCentralWp}{2312}
\newcommand{\AveTotStatWp}{  26}
\newcommand{\AveTotSystWp}{  27}
\newcommand{\AveTotExtrapWp}{  30}
\newcommand{\AveTotLumiWp}{  72}

\newcommand{\RatCombCentWmoverZ}{3.9147}
\newcommand{\RatCombStatWmoverZ}{0.1330}
\newcommand{\RatCombSystWmoverZ}{0.0376}
\newcommand{\RatCombCentWpoverZ}{7.0343}
\newcommand{\RatCombStatWpoverZ}{0.2279}
\newcommand{\RatCombSystWpoverZ}{0.0646}
\newcommand{\RatCombCentWpoverWm}{1.797}
\newcommand{\RatCombStatWpoverWm}{0.034}
\newcommand{\RatCombSystWpoverWm}{0.009}
\newcommand{\AveTotCentralW}{3711}
\newcommand{\AveTotStatW}{  34}
\newcommand{\AveTotSystW}{  43}
\newcommand{\AveTotExtrapW}{  51}
\newcommand{\AveTotLumiW}{ 115}
\newcommand{\AveFidCentralW}{2231}
\newcommand{\AveFidStatW}{  20}
\newcommand{\AveFidSystW}{  26}
\newcommand{\AveFidLumiW}{  69}
\usepackage{ANA-STDM-2018-06-PAPER-defs}
\usepackage{my_ratios}
 
\graphicspath{{logos/}{figures/}}
 
\usepackage{multirow}

\AtlasTitle{Measurement of $W^{\pm}$-boson and $Z$-boson production cross-sections in $pp$ collisions at $\sqrt{s}=2.76$~TeV with the ATLAS detector}

\AtlasAbstract{
The production cross-sections for $W^{\pm}$ and $Z$ bosons are
measured using ATLAS data corresponding to an integrated luminosity
of 4.0~pb$^{-1}$  collected at a centre-of-mass energy
$\sqrt{s}=2.76$~TeV. The decay channels $W \rightarrow \ell \nu$
and $Z \rightarrow \ell \ell $ are used, where $\ell$ can be an
electron or a muon. The cross-sections are presented for a fiducial
region defined by the detector acceptance and are also extrapolated
to the full phase space for the total inclusive production cross-section. The combined (average) total inclusive cross-sections for the electron and muon channels are:

\begin{eqnarray}
\sigma^{\text{tot}}_{W^{+}\rightarrow \ell \nu}& =  & 2312  \pm 26\ (\text{stat.})\ \pm 27\ (\text{syst.}) \pm 72\ (\text{lumi.})  \pm 30\  (\text{extr.})~\text{pb} \nonumber, \\
\sigma^{\text{tot}}_{W^{-}\rightarrow \ell \nu}& =  & 1399  \pm 21\ (\text{stat.})\ \pm 17\ (\text{syst.}) \pm 43\ (\text{lumi.}) \pm 21\ (\text{extr.})~\text{pb} \nonumber, \\
\sigma^{\text{tot}}_{Z \rightarrow \ell \ell}& =  &  323.4  \pm 9.8\
(\text{stat.}) \pm 5.0\ (\text{syst.}) \pm 10.0\ (\text{lumi.})  \pm 5.5 (\text{extr.}) ~\text{pb} \nonumber.
\end{eqnarray}
 
Measured ratios and asymmetries constructed using these cross-sections are also presented. These observables benefit from full or partial cancellation of many systematic uncertainties that are correlated between the different measurements.
 
}

\AtlasRefCode{STDM-2018-06}
 
\PreprintIdNumber{CERN-EP-2019-095}

\LEcontact{Steve Lloyd s.l.lloyd@qmul.ac.uk}
\HepDataRecord{https://www.hepdata.net/record/ins1742785}
 
\AtlasJournal{Eur. Phys. J. C}
\AtlasJournalRef{Eur. Phys. J. C 79 (2019) 901}
\AtlasDOI{10.1140/epjc/s10052-019-7399-7}

\AtlasCoverSupportingNote{Support Note}{https://cds.cern.ch/record/2296490}
 
\AtlasCoverCommentsDeadline{1 July 2019}
 
\AtlasCoverAnalysisTeam{
F.~Balli, M.~Boonekamp, C.~Debenedetti, J.~Ferrando, K.~Gasnikova,
A.~Glazov, U.~Klein, K.~Lohwasser, E~Lobodzinska, E.~Rizvi,
K.~Schmieden,F.~Sforza, G.~Siragusa, A.~Trofymov, M.~Vincter, N.~Zakharchuk}
 
\AtlasCoverEdBoardMember{P.~Krieger~(chair), D.~Fassouliotis, C.~Sawyer}
 
\AtlasCoverEgroupEditors{atlas-stdm-2018-06-contact-editors@cern.ch}
 
\AtlasCoverEgroupEdBoard{atlas-stdm-2018-06-editorial-board@cern.ch}
 
\hypersetup{pdftitle={ATLAS document},pdfauthor={The ATLAS Collaboration}}
 
\begin{document}
 
\maketitle
 
\tableofcontents

% The next lines are included from the .//tex/Introduction.tex input file
\section{Introduction}
\label{sec:intro}
 
The processes that  produce $W$ and $Z$ bosons\footnote{In this paper it is implicit that $Z$ boson refers to $Z/\gamma^*$ bosons.}  in $pp$ collisions via Drell--Yan
annihilation are two of the simplest at hadron colliders to describe theoretically.  At lowest
order in quantum chromodynamics (QCD), $W$-boson production proceeds via $q\bar{q}' \rightarrow W$ and $Z$-boson
production via $q\bar{q} \rightarrow Z$. Therefore, precision measurements of these
production cross-sections yield
important information about the  parton distribution functions (PDFs) for
quarks inside the proton. Factorisation theory allows PDFs to be treated separately from the perturbative QCD high-scale collision calculation as  functions of the event  energy  scale, $Q$, and the momentum fraction of the parton, $x$, for each parton flavour.
Usually PDFs are defined for a particular starting scale $Q_0$ and can be
evolved to other scales via the DGLAP equations~\cite{Gribov:1972ri,Lipatov:1974qm,Altarelli:1977zs,Dokshitzer:1977sg}.
Measurements of on-shell $W/Z$-boson production probe the PDFs in a
range of $Q^2$ that lies  close to $m^2_{W/Z}$. The range of $x$ that is probed depends
on the centre-of-mass energy, $\sqrt{s}$, of the protons and the rapidity coverage of the detector. Each measurement of these production cross-sections at a new value of $\sqrt{s}$ thus provides information complementary to previous measurements. The combinations of initial partons participating in the production
processes of $W^+$,$W^-$, and $Z$ bosons are different, so
each process provides complementary information about the products of different quark PDFs.
 
This paper presents the first measurements of the production cross-sections for $W^+$, $W^-$ and $Z$ bosons in $pp$ collisions at $\sqrt{s}=2.76$~\TeV{}. The data
were collected by the ATLAS detector at the Large Hadron Collider (LHC)~\cite{Evans:2008zzb} in 2013 and correspond
to an integrated luminosity of 4.0~pb$^{-1}$. To provide further sensitivity to PDFs,
and to reduce the systematic uncertainty in the predictions, ratios of these cross-sections and the charge asymmetry for $W$-boson production are also presented.
The measurements are performed for leptonic (electron or muon) decays of the $W$
and $Z$ bosons, in a defined fiducial region, and also extrapolated to the
total cross-section.
 
Previous measurements of the $W$-boson and $Z$-boson production cross-sections in $pp$
collisons at the  LHC were performed by the ATLAS, CMS and LHCb Collaborations at   $\sqrt{s}=5.02$~\TeV{}~\cite{Aaboud:2018nic}, 7~\TeV{}~\cite{Aaboud:2016btc,CMS:2011aa,Aaij:2015gna,Aaij:2014wba,Aaij:2012mda}, 8~\TeV{}~\cite{Aad:2015auj,Chatrchyan:2014mua,Aaij:2015zlq,Aaij:2016qqz,Aaij:2015vua} and
13~\TeV{}~\cite{Aad:2016naf,Aaboud:2016zpd,Aaij:2016mgv}, and by the PHENIX and STAR Collaborations at the RHIC at $\sqrt{s}=500$~\GeV{}~\cite{Adare:2010xa,STAR:2011aa,} and 510~\GeV{}~\cite{Adare:2018csm}. This is the first measurement at 2.76~\TeV{}. Other measurements
of these processes were performed in $p\bar{p}$ collisons at $\sqrt{s} = 1.8$~\TeV{} and 1.96~\TeV{} by the CDF~\cite{Abe:1992ys,Abe:1995bm,Abe:1998gq,Affolder:1999jh,Abulencia:2005ix} and D0~\cite{Abachi:1995xc} Collaborations, and at $\sqrt{s}=546$~\GeV{} and $630$~\GeV{} by the UA1~\cite{Albajar:1988ka} and UA2~\cite{Alitti:1991dm} Collaborations.

% End of text imported from the .//tex/Introduction.tex input file
 
% The next lines are included from the .//tex/ATLAS.tex input file
\section{ATLAS detector}
\label{sec:detector}
 
The ATLAS detector~\cite{PERF-2007-01} at the LHC covers nearly the entire solid angle around the collision point.
It consists of an inner tracking detector surrounded by a thin superconducting solenoid, electromagnetic (EM) and hadronic calorimeters,
and a muon spectrometer (MS) incorporating three large superconducting toroid magnets.
The inner-detector system (ID) is immersed in a \SI{2}{\tesla} axial magnetic field
and provides charged-particle tracking in the pseudorapidity range $|\eta| < 2.5$.\footnote{ATLAS uses a right-handed coordinate system with its origin at the nominal interaction point (IP)
in the centre of the detector and the $z$-axis along the beam pipe.
The $x$-axis points from the IP to the centre of the LHC ring,
and the $y$-axis points upwards.
Cylindrical coordinates $(r,\phi)$ are used in the transverse plane,
$\phi$ being the azimuthal angle around the $z$-axis.
The pseudorapidity is defined in terms of the polar angle $\theta$ as $\eta = -\ln \tan(\theta/2)$.
Angular distance is measured in units of $\Delta R \equiv \sqrt{(\Delta\eta)^{2} + (\Delta\phi)^{2}}$.}
 
The high-granularity silicon pixel detector covers the vertex region and typically provides three measurements per track.
It is followed by the silicon microstrip tracker, which usually
provides eight measurements from eight strip layers.
These silicon detectors are complemented by the transition radiation tracker (TRT),
which enables radially extended track reconstruction up to $|\eta| = 2.0$.
The TRT also provides electron identification information
based on the fraction of hits (typically 30 in total) above a higher energy-deposit threshold associated with the presence of transition radiation.
 
The calorimeter system covers the pseudorapidity range $|\eta| < 4.9$.
Within the region $|\eta|< 3.2$, EM calorimetry is provided by barrel and
endcap high-granularity lead/liquid-argon (LAr) sampling calorimeters,
with an additional thin LAr presampler covering $|\eta| < 1.8$
that is used to correct for energy loss in material upstream of the calorimeters.
Hadronic calorimetry in this region is provided by the steel/scintillator-tile calorimeter,
segmented into three barrel structures with $|\eta| < 1.7$, and two copper/LAr hadronic endcap calorimeters.
The solid angle coverage is completed with forward copper/LAr and tungsten/LAr calorimeter modules
optimised for EM and hadronic measurements, respectively.
 
The muon spectrometer comprises separate trigger and
high-precision tracking chambers measuring the deflection of muons in a magnetic field generated by superconducting air-core toroids.
The precision chamber system covers the region $|\eta| < 2.7$ with three layers of monitored drift tubes,
complemented by cathode strip chambers in the forward region, where the backgrounds are highest.
The muon trigger system covers the range $|\eta| < 2.4$ with resistive plate chambers in the barrel and thin gap chambers in the endcap regions.

The ATLAS detector selected events using a three-level trigger system~\cite{PERF-2011-02}.
The first-level trigger is implemented in hardware and used a subset of detector information
to reduce the event rate to a design value of at most \SI{75}{\kHz}.
This was followed by two software-based triggers that together reduced the event rate to about \SI{200}{\Hz}.

% End of text imported from the .//tex/ATLAS.tex input file
 
% The next lines are included from the .//tex/Data.tex input file
\section{Data and simulation samples}
\label{sec:samples}
 
The data used in this measurement were collected in February 2013 during a period when proton beams at the LHC were collided at a centre-of-mass energy of 2.76 \TeV{}. During this running period a typical value of the instantaneous luminosity was $1 \times 10^{32}$\ cm$^{-2}$s$^{-1}$, significantly lower than in 7, 8 and 13~\TeV{} data-taking conditions. The typical value of the  mean number of collisions per  proton bunch crossing (pile-up) $\langle \mu \rangle$ was  0.3. Only data from stable collisions when the ATLAS detector was fully operational are used, yielding a data sample corresponding to an integrated luminosity of 4.0\ pb$^{-1}$.
 
Samples of Monte Carlo (MC) simulated events are used to estimate the signals from $W$-boson and $Z$-boson production, and the backgrounds containing prompt leptons: electroweak-diboson production and top-quark pair ($t\bar{t}$) production. Background  contributions arising from  multijet events that do not contain prompt leptons  are estimated directly from data, with simulated events used to cross-check these estimations in the muon channel.

Production of single $W$ and $Z$ bosons was simulated using \POWHEGBOXV{v1 r1556}~\cite{Nason:2004rx,Frixione:2007vw,Alioli:2010xd,Alioli:2008gx}. The parton showering was performed using \PYTHIAV{8.17}~\cite{Sjostrand:2007gs}. The PDF set used for the simulation was CT10~\cite{Lai:2010vv}, and the parton shower parameter values were those of the AU2 tune~\cite{ATL-PHYS-PUB-2014-021}. Additional quantum electrodynamics (QED) emissions from electroweak (EW) vertices and charged leptons were simulated using \textsc{Photos++} v3.52~\cite{Davidson:2010ew}. Additional samples of simulated $W$-boson events generated with \SHERPAV{2.1}~\cite{Gleisberg:2008ta} are used to estimate uncertainties arising from the choice of event generator model. In these  \SHERPA\ samples, simulation of $W$-boson production in association with up to two additional partons was performed at next-to-leading order (NLO) in QCD while production of $W$ bosons in association with three or four additional partons was performed at leading order (LO) in QCD\@. The sample cross-sections  were normalised to next-to-next-to-leading-order (NNLO) QCD predictions for the total cross-sections described in Section~\ref{sec:results}.
 
\POWHEGBOXV{v1 r2330} was used to generate $t\bar{t}$ samples~\cite{Frixione:2007nw}. These samples had parton showering performed using \PYTHIAV{6.428}~\cite{Sjostrand:2006za} with parameters corresponding to the Perugia2011C tune~\cite{Skands:2010ak}. The CT10 PDF set was used. Additional QED final-state radiative corrections were applied using \textsc{Photos++} v3.52 and $\tau$-lepton decays were performed using \textsc{Tauola}  v25feb06~\cite{Davidson:2010rw}. Single production of top quarks is a negligible contribution to this analysis, compared with $t\bar{t}$ production, so no such samples were generated.
 
Production of two massive electroweak bosons  ($WW, ZZ, WZ$) was simulated using \HERWIGV{6.5}~\cite{Corcella:2002jc}, with multiparton interactions modelled using \JIMMYV{4.13}~\cite{Butterworth:1996zw}. The CTEQ6L1 PDF set~\cite{Pumplin:2002vw} and AUET2 tune~\cite{ATLAS:2011gmi} were used for these samples.
 
Multijet production containing heavy-flavour final states, arising from the production of $b\bar{b}$ or $c\bar{c}$ pairs, were simulated using \PYTHIAV{8.186}. The CTEQ6L1 PDF set and AU2 tune were used. Events were required to contain an electron or muon with transverse momentum $p_{\text{T}} > 10$~\GeV{} and $|\eta| <2.8$.
 
The detector response to generated events was simulated by passing the events through a model of the ATLAS detector~\cite{SOFT-2010-01} based on \GEANT{}4~\cite{Agostinelli:2002hh}. Additional minimum-bias events generated using \PYTHIAV{8.17} and the A2 set of tuned parameters,  were overlaid in such a way that the distribution of $\langle \mu \rangle$ for simulated events reproduced that in the real data. The resulting events were then passed through the same reconstruction software as the real data.
 
The simulated samples used for the baseline analysis are summarised in Table~\ref{tab:mc}, which shows the generator used for each process together with the order in QCD at which they were generated.
 
\begin{table}
\caption{Summary of the baseline simulated samples used\label{tab:mc}.}
\begin{center}
\begin{tabular}{l|l|c}
Process & Generator & Generator QCD precision   \\
\hline
\multicolumn{3}{c}{Signal Samples}\\
\hline
$W \rightarrow \ell \nu $ & \POWHEGBOX+\PYTHIAV{8} & NLO \\
$Z \rightarrow \ell^+ \ell^- $ & \POWHEGBOX+\PYTHIAV{8} & NLO\\
\hline
\multicolumn{3}{c}{Background Samples}\\
\hline
$W \rightarrow \tau \nu$ & \POWHEGBOX+\PYTHIAV{8} & NLO \\
$Z \rightarrow \tau^+ \tau^- $ & \POWHEGBOX+\PYTHIAV{8} & NLO\\
$t\bar{t}$ & \POWHEGBOX+\PYTHIAV{6} & NLO \\
$WW$ & \HERWIG\ & LO \\
$ZZ$ & \HERWIG\ & LO \\
$WZ$ & \HERWIG\ & LO \\
$b\bar{b}$ & \PYTHIAV{8} & LO \\
$c\bar{c}$ & \PYTHIAV{8} & LO \\
\end{tabular}
\end{center}
\end{table}
% End of text imported from the .//tex/Data.tex input file
 
% The next lines are included from the .//tex/Selection.tex input file
\section{Event selection}
\label{sec:selection}
 
This section describes the selection of events consistent with the production
of $W$ bosons or $Z$ bosons.  The $W$-boson selection requires events to contain a single charged lepton and large missing transverse momentum. The $Z$-boson selection requires events to contain two charged leptons with opposite charge and the same flavour.

Events were selected by triggers
that required at least one charged electron (muon) with
$p_{\text{T}} > 15~$\GeV{} (10~\GeV{}). These thresholds  yield an event sample with a uniform efficiency as a function of the $E_{\text{T}}$ and $p_{\text{T}}$ requirements used subsequently to select the final event sample. The hard-scatter vertex, defined as the vertex with highest sum of squared track transverse momenta (for tracks with $p_{\text{T}} > 400$~\MeV{}), is required to have at least three associated tracks.

Electrons are reconstructed from clusters of energy in the EM calorimeter that are
matched to a track reconstructed in the ID. The electron is
required to have $p_{\text{T}} > 20$~\GeV{} and $|\eta| < 2.4$ (excluding the
transition region between barrel and endcap calorimeters of
$1.37 < |\eta| < 1.52$). Each electron must satisfy a set of identification criteria  designed
to suppress misidentified photons or jets. Electrons are required to satisfy the
\textit{medium} selection, following the definition provided in
Ref. ~\cite{Aaboud:2016vfy}. This includes requirements on the shower shape in
the EM calorimeter, the leakage of the shower into the hadronic
calorimeter, the number of hits measured along the track in the ID,
and the quality of the cluster–track matching. A Gaussian sum filter~\cite{ATLAS-CONF-2012-047} algorithm is used to re-fit the tracks and improve the estimated electron track parameters. 
To suppress background from misidentified objects such as jets, the electron is required to be isolated using calorimeter-based criteria. The sum of the transverse energies of clusters lying within a cone of size  $\Delta R = 0.2$ around the centroid of the electron cluster  and excluding the core\footnote{The core of the shower is the contribution within $\Delta \eta \times \Delta
\phi = 0.125 \times 0.175$ around the cluster barycentre.} must be less than 10\% of the electron $p_{\text{T}}$.

Muon candidates are reconstructed by combining tracks reconstructed in  the ID with tracks reconstructed in the MS~\cite{Aad:2014rra}. They are required to have $p_{\text{T}} > 20$~\GeV{} and $| \eta | < 2.4$. 
The muon candidates are also required to be isolated, by requiring that the scalar sum of the $p_{\text{T}}$ of additional tracks within a cone of size $\Delta R = 0.4$ around the muon is less than 80\% of the muon $p_{\text{T}}$.

The missing transverse momentum vector~\cite{Aad:2016nrq}  ($\boldsymbol{E^{\text{miss}}_{\text{T}}}$) is calculated as the negative vector sum of the transverse momenta of electrons and muons, and of the transverse momentum of the recoil. The magnitude of this vector is denoted by $E_{\text{T}}^{\text{miss}}$. The recoil vector is obtained by summing the transverse momenta of all clusters of energy  measured in the calorimeter, excluding those within  $\Delta R = 0.2$ of the lepton candidate. The momentum vector of each cluster is determined by the magnitude and coordinates of the energy deposits; the cluster is assumed to be massless. Cluster energies are initially measured assuming that the energy deposition occurs only through EM interactions, and are then corrected for the different calorimeter responses to hadrons and electromagnetically interacting particles, for losses due to dead material, and for energy that is not captured by the clustering process~\cite{PERF-2014-07}. The definition of the recoil does not make use of reconstructed jets, to avoid threshold effects. The procedure used to calibrate the recoil closely follows that used in the recent ATLAS measurement of the $W$-boson mass~\cite{Aaboud:2017svj}, first correcting the modelling of the overall recoil in simulation and then applying corrections for residual differences in the recoil response and resolution that are derived from $Z$-boson data and transferred to the $W$-boson sample.

The $W$-boson selection requires events to contain exactly one lepton (electron or muon) candidate and  have $E_{\text{T}}^{\text{miss}}> 25$~\GeV{}. The lepton must match a lepton candidate that met the trigger
criteria.  The transverse  mass, $m_{\text{T}}$, of the $W$-boson candidate in the event is calculated using the lepton candidate and $E_{\text{T}}^{\text{miss}}$ according to
$ m_{\text{T}} = \sqrt{2p^{\ell}_{\text{T}} E_{\text{T}}^{\text{miss}}  (1 - \cos (\phi_\ell - \phi_{E_{\text{T}}^{\text{miss}}}))}$. The transverse mass in $W$-boson production events is expected to exhibit a Jacobian peak around the $W$-boson mass. Thus, requiring that $m_{\text{T}}>40~\GeV{}$ suppresses background processes. After these requirements there are 3914 events in the $W \rightarrow e^+ \nu$ channel, 2209  events in the $W \rightarrow e^- \bar{\nu}$ channel, 4365  events in the $W \rightarrow \mu^+ \nu$ channel, and 2460 events in the $W \rightarrow \mu^- \bar{\nu}$ channel.

The $Z$-boson selection requires events to contain exactly two lepton candidates with the same flavour and opposite charge. At least one lepton must match a lepton candidate that met the trigger
criteria. Background processes  are suppressed by requiring that the invariant mass of the lepton pair satisfies $66 < m_{\ell \ell} < 116$~\GeV{}. After these requirements there are 430 events in the $Z\rightarrow e^+ e^-$ channel, and 646 events in the $Z\rightarrow \mu^+ \mu^-$ channel.

% End of text imported from the .//tex/Selection.tex input file
 
% The next lines are included from the .//tex/Background.tex input file
\section{Background estimation}
\label{sec:backgrounds}
 
The background processes that contribute to the sample of events passing the $W$-boson and $Z$-boson selections can be separated into two categories: those estimated from MC simulation and theoretical calculations, and those estimated directly from data. The main backgrounds that contribute to  the event sample passing the $W$-boson selection are processes with a $\tau$-lepton decaying into an electron or muon plus neutrinos, leptonic $Z$-boson decays where only one lepton is reconstructed, and multijet processes. The main background contribution to the event sample passing the $Z$-boson selection is production of two massive electroweak bosons.

The backgrounds arising from $W \rightarrow \tau \nu$,  $Z\rightarrow \ell^+ \ell^-$, diboson production, and $t\bar{t}$ production are estimated from the simulated samples described in Section~\ref{sec:samples}.
Predictions of the backgrounds to the $W$-boson and $Z$-boson production
measurements arising from multijet production suffer from large theoretical
uncertainties, and therefore the contribution to  this background in the $W$-boson measurement is
estimated from data. This is achieved by  constructing a shape template for the background using a discriminating variable in a control region and then performing a template fit to the same distribution in the signal region to extract the background contribution. The choice of  template variable is motivated by the difference between signal and background and by the available number of events.
Previous ATLAS measurements at 7~\TeV{}~\cite{Aaboud:2016btc} and 13~\TeV{}~\cite{Aaboud:2016zpd} found that  multijet production  makes a  background contribution of less than 0.1\% for $Z$-boson measurements; this is therefore neglected.

Electron candidates in multijet background events are typically misidentified candidates produced when jets mimic the signature of an electron, for example when a neutral pion and a charged pion overlap in the detector. Additional candidates can arise from 'non-prompt' electrons produced when a photon converts, and in decays of heavy-flavour hadrons.
To construct a control region for the multijet template, a selection is used that differs from the $W$-boson selection described in Section~\ref{sec:selection} in only two respects: the medium electron identification criteria are inverted (while keeping the looser identification criteria) and the $E_{\text{T}}^{\text{miss}}$ requirement is removed. By construction, this control region is statistically independent of the $W$-boson signal region. A template for the shape of the multijet background in the  $E_{\text{T}}^{\text{miss}}$ distribution is then obtained from that distribution in the control region after subtraction of expected contributions from the signal and other backgrounds determined using MC samples.
The normalisation of the multijet background template in the signal region is extracted by performing a $\chi^2$ fit of the $E_{\text{T}}^{\text{miss}}$ distribution (applying all signal criteria except the requirement on $E_{\text{T}}^{\text{miss}}$) to a sum of the templates for the multijet background, the signal, and all other backgrounds. The normalisation of the signal is allowed to vary freely in the fit as is the  multijet background; however, the other backgrounds are only allowed to vary from their expected values by up to 5\%, corresponding to the largest level of variation in predicted electroweak-boson production cross-sections obtained from varying the choice of PDF. The normalisation from this fit can then be used together with the inverted selection to construct multijet background distributions in any other variable that is not correlated with the electron identification criteria.

Muon candidates in multijet background events are typically `non-prompt' muons produced in the decays of hadrons. The multijet background contribution to the $W \rightarrow \mu \nu$ selection
is estimated by using the same method as described for the $W \rightarrow e \nu $
selection. In this case the control region is defined by inverting the isolation requirement and removing the requirement on $m_{\text{T}}$. The distribution used for the fits is $m_{\text{T}}$.

\begin{figure}[!tbp]
\centering
\subfloat[]{\includegraphics[width=0.8 \textwidth]{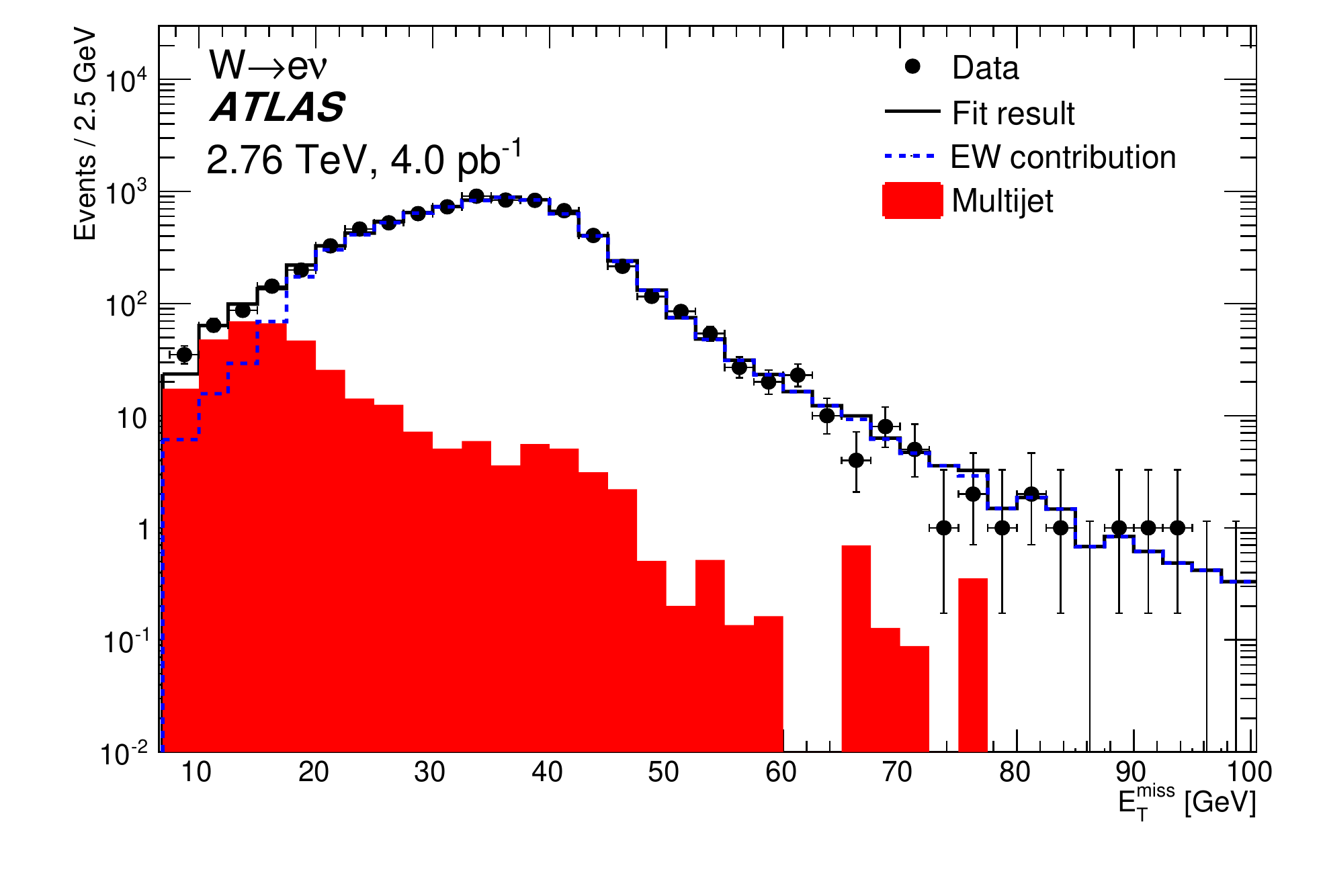}}\\
\subfloat[]{\includegraphics[width=0.8 \textwidth]{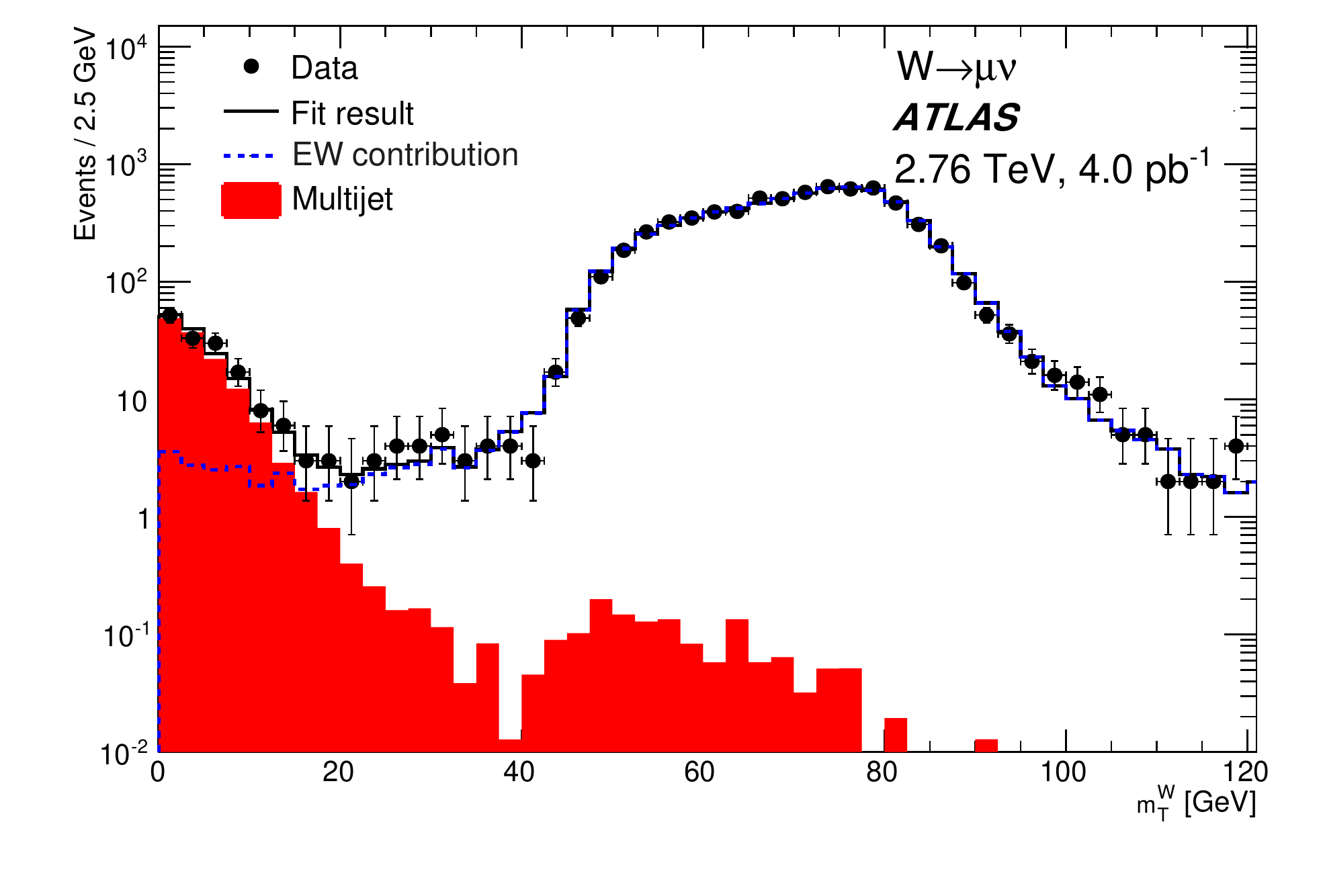}}
\caption{ Distributions used to estimate the multijet background
contribution in (a) the $W \rightarrow e \nu$ channel, and (b) the
$W \rightarrow \mu \nu$ channel.  The data is compared to the fit result. \label{fig:mjback}}
\end{figure}

The overall number of multijet background events is estimated from a fit to the total $W$-boson sample. Comparisons between the fitted distributions and data for $W\rightarrow e\nu$ and $W\rightarrow \mu \nu$ are shown in Figure~\ref{fig:mjback}. Fits to the separate $W^+$-boson and $W^-$-boson samples are used in the  evaluation of the systematic uncertainties, as described in Section~\ref{sec:systematics}.
The final estimated multijet contributions are $30 \pm 11$  events for $W \rightarrow e^+ \nu$ and $W \rightarrow e^- \nu$ and $2.5 \pm 1.9$ events  for  $W^+ \rightarrow \mu^+\nu$ and $W^-\rightarrow \mu^- \nu$. The relative contribution of the multijet events (1\%) is lower than in 13~\TeV\ (4\%) and 7~\TeV\ (3\%) data. This is in agreement with  expectations for this lower pile-up running, where the
resolution in $E^{\text{miss}}_{\text{T}}$ is improved compared to the higher pile-up running.
 
% End of text imported from the .//tex/Background.tex input file
 
% The next lines are included from the .//tex/Unfolding.tex input file
\section{Correction for detector effects}
\label{sec:unfolding}
 
The measurements in this paper are performed within specific fiducial regions
and extrapolated to the total $W$-boson or $Z$-boson phase space. The fiducial
regions are defined by the kinematic and geometric selection criteria
given in Table~\ref{tab:fiducial}; in simulations these are applied  at the generator level before  the emission of QED final-state radiation from the decay lepton(s) (QED Born  level).

The  fiducial $W$-boson/$Z$-boson  production cross-section is obtained from the number of observed events meeting the selection criteria after background contributions are subtracted, $N^{\text{sig}}_{W,Z} $,
using the following formula:

\begin{equation*}
\sigma^{\text{fid}}_{W,Z  \rightarrow \ell \nu, \ell \ell} = \frac{ N^{\text{sig}}_{W,Z} }{ C_{W,Z}\cdot \mathcal{L}_{\text{int}}},
\end{equation*}
 
where $ \mathcal{L}_{\text{int}}$ is the total integrated luminosity of the data
samples used for the analysis.
The factor $C_{W,Z}$ is the ratio of the number of generated events that satisfy the final
selection criteria after event reconstruction to the  number of
generated events within the fiducial region. It includes the efficiency for
triggering,  reconstruction and identification of $W,Z \rightarrow \ell \nu, \ell^+ \ell^-$ events falling  within the acceptance.
The different components of
the efficiency are calculated using a mixture of MC simulation and
measurements from data.
 
The  total $W$-boson and $Z-$boson production cross-sections are obtained using the following formula:

\begin{equation*}
\sigma^{\text{tot}}_{W,Z  \rightarrow \ell \nu, \ell \ell} \equiv
\sigma^{\text{tot}}  \times {B}(W,Z  \rightarrow \ell \nu, \ell \ell) = \frac{ N^{\text{sig}}_{W,Z} }{A_{W,Z} \cdot C_{W,Z}\cdot \mathcal{L}_{\text{int}}}.
\end{equation*}
 
The factor  ${B}(W,Z  \rightarrow \ell \nu, \ell \ell)$
is  the per-lepton branching fraction of the vector boson. The factor $A_{W,Z}$ is the acceptance for
$W/Z$-boson events being studied.  It is  defined as the fraction of
generated events that  satisfy the fiducial requirements.
This acceptance is  determined using MC signal samples,
corrected to the generator QED Born level, and is used to extrapolate
the  measured cross-section in the fiducial region to the full phase space.
The central values of $A_{W,Z}$ are around 0.6 for these measurements,
compared with 0.5 at $\sqrt{s}=7$~\TeV\ and  0.4 at $\sqrt{s}=13$~\TeV,
so  the fiducial region is closer to the full phase space in this
measurement than for those at higher centre-of-mass energies. This is
due to a combination of higher $p_{\text{T}}$ thresholds for
leptons in other measurements, and more-central production of
vector bosons at lower $\sqrt{s}$. The values of $C_W$ are
approximately 0.67 for the $W \rightarrow e\nu$ channels and 0.75
for the $W \rightarrow \mu \nu$ channels. The values of $C_Z$ are
0.55 for the $Z\rightarrow e^+ e^-$ channel and 0.79 for $Z
\rightarrow \mu^+ \mu^-$. The $C_{W,Z}$ values are a little higher
than for previous measurements at $\sqrt{s}=7$~\TeV{} and $\sqrt{s}=13$~\TeV{}.
 
\begin{table}
\caption{Summary of the selection criteria that define the measured fiducial regions.\label{tab:fiducial}}
\begin{center}
\begin{tabular}{c|c}
$W$-boson fiducial region & $Z$-boson fiducial region \\
\hline
\hspace{0.5cm}$p^{\ell}_{\text{T}} > 20$~\GeV{} \hspace{0.5cm} & $p^{\ell^{+,-}}_{\text{T}} > 20$~\GeV{}\\
|$\eta^{\ell}| < 2.4$\hspace{0.9cm} & $|\eta^{\ell^{+,-}}| < 2.4$\hspace{0.9cm} \\
$E^{\text{miss}}_{\text{T}} > 25$~\GeV{}\hspace{0.45cm} & $66< m_{\ell^+ \ell^-} < 116$~\GeV{} \\
$m_{\text{T}}> 40$~\GeV{}\hspace{0.1cm}    &  \\
\end{tabular}
\end{center}

\end{table}
% End of text imported from the .//tex/Unfolding.tex input file
 
% The next lines are included from the .//tex/Systematics.tex input file
\section{Systematic uncertainties}
\label{sec:systematics}
 
The systematic uncertainty in the electron
reconstruction and identification efficiency is estimated using the tag-and-probe method in
8~\TeV{} data~\cite{Aaboud:2016vfy,Aaboud:2017soa}  and extrapolated
to the 2.76~\TeV{} dataset. The extrapolation procedure results in absolute increases of $\pm$2\%, due
to uncertainties in the effect of the differing pile-up conditions in
2.76~\TeV{} data relative to the 8~\TeV{}
data, as well as a different setting of the noise filtering in the LAr
calorimeter of the 2.76 TeV data relative to the 8 TeV data.
These uncertainties were estimated  using a comparison between 7 TeV and
8 TeV data and MC samples, after having established that the central
values of the efficiencies are the same for different centre-of-mass
energies when the same LAr filter settings are used. A similar
methodology had been used for internal estimates of the electron
efficiency performance at 13 TeV before the start of
Run-2 data taking and was found to give a good prediction of the
efficiencies in data as well as a conservative estimate of the uncertainties.
Transverse-momentum-dependent isolation corrections, calculated
with the tag-and-probe method in 2.76~\TeV{} data, are very close to 1, so the systematic
uncertainty in the electron isolation requirement is set to the size
of the correction itself, that is  $\pm1$\% for low $p_{\text{T}}$ and $\pm0.3$\% for higher $p_{\text{T}}$.
The electron energy scale has  associated statistical uncertainties and systematic uncertainties arising from a
possible bias in the
calibration method, the choice of generator,  the
presampler energy scale, and imperfect knowledge of the material in front of
the EM calorimeter~\cite{PERF-2010-04}. The total energy-scale
uncertainty is calculated as the sum in quadrature of these components.

Systematic uncertainties associated with the  muon momentum  can be divided
into three major independent categories: momentum resolution of the MS
track, momentum resolution of the ID track, and an overall scale uncertainty~\cite{Aad:2014rra}.
The total momentum scale/resolution uncertainty is the sum in quadrature of these components.
An $\eta$-independent uncertainty of approximately $\pm1.1$\% in the muon trigger efficiency, determined using the
tag-and-probe method~\cite{Aad:2014rra} in  2.76~\TeV{} data, is taken into account.  Furthermore, a $p_{\text{T}}$- and $\eta$- dependent uncertainty in
the identification and reconstruction efficiencies of approximately $\pm0.3$\,\%, derived using the
tag-and-probe method on 8\,\TeV{} data is applied. The uncertainty in
the $p_{\text{T}}$-dependent isolation correction in the muon channel,
calculated with the tag-and-probe method in 2.76~\TeV{} data, is about $\pm0.6$\% for low $p_{\text{T}}$ and $\pm0.5$\% for higher $p_{\text{T}}$.
 
The luminosity uncertainty for the 2.76~\TeV\ data is $\pm3.1$\%. This is determined, following the same methodology as was used for the 7~\TeV{} data  recorded in 2011~\cite{DAPR-2011-01}, from a calibration of the luminosity scale derived from beam-separation scans performed during the 2.76~\TeV\ operation of the LHC in 2013.

Systematic uncertainties in the $E^{\text{miss}}_{\text{T}}$ arising
from the smearing and bias corrections applied to obtain satisfactory
modelling of the recoil~\cite{Aad:2016nrq}
affect the $C_W$ factors in the $W \rightarrow \ell \nu$ measurement,
and are taken into account.

Uncertainties arising from the choice of PDF set are evaluated using
the error sets of the initial CT10  PDF set (at 90\% confidence level (CL)) and from
comparison with the results obtained using the central PDF sets from
ABKM09~\cite{Alekhin:2009ni}, NNPDF23~\cite{Ball:2012cx}, and
ATLAS-epWZ12~\cite{Aad:2012sb}. The effect of this uncertainty on
$A_{W^+}$ ($A_{W^-}$) is estimated to be $\pm1.0$\% (1.2\%), and the effect
on $A_Z$ is estimated to be $\pm$1.4\%. The effect on $C_{W,Z}$ is between
$\pm0.05$\% and $\pm0.4$\% depending on the channel.
 
\begin{table}
\caption{Relative systematic uncertainties (\%) in the correction factors $C_{W,Z}$ in different channels.\label{tab:systs}}
\label{tab:Unc}
\begin{center}
\begin{small}
\begin{tabular}{l | c   c  c | c  c  c   }
$\delta C / C [\%]$ & $W^{\!+}\!\!\to\!\!e^{\!+}\nu$& $W^{\!-}\!\!\to\!\!e^{\!-}\nu$& $Z\!\!\to\!\!e^{\!+}e^{\!-}$ & $W^{\!+}\!\!\to\!\!\mu^{\!+}\nu$ & $W^{\!-}\!\!\!\to\!\!\mu^{\!-}\nu$ & $Z\!\!\to\!\!\mu^{\!+}\mu^{\!-}$ \\
\hline
Lepton trigger & 0.14 & 0.13 & $<0.01$ & 1.07 & 1.07 & 0.03  \\
Lepton reconstr. and ident. & 2.31 & 2.33 & \hspace{0.3cm}4.55 & 0.30 & 0.32 & 0.62 \\
Lepton isolation & 0.71 & 0.71 & \hspace{0.3cm}1.41 & 0.51 & 0.51 & 1.01 \\
Lepton scale and resolution & 0.44 & 0.43 & \hspace{0.3cm}0.34 & 0.05 & 0.05 & 0.04 \\
Recoil scale and resolution & 0.25 & 0.20 & -- & 0.22 & 0.22 & -- \\
PDF & 0.22 & 0.29 & \hspace{0.3cm}0.11 & 0.11 & 0.20 & 0.06 \\
MC statistical uncertainty & 0.24 & 0.31 & \hspace{0.3cm}0.30 & 0.24 & 0.34 & 0.43 \\
\hline
Total & 2.5\hspace{0.15cm} & 2.5\hspace{0.15cm} & \hspace{0.15cm}4.8 & 1.3\hspace{0.15cm} & 1.3\hspace{0.15cm} & 1.3\hspace{0.15cm} \\
\end{tabular}
\end{small}
\end{center}
\end{table}
 
A summary of the systematic uncertainties in the $C_{W,Z}$ factors is
shown in Table~\ref{tab:systs}. The muon trigger, and electron
reconstruction and identification uncertainties are dominant.
 
Uncertainties arising from the choice of event generator and parton shower
models are estimated by comparing results obtained when using
\SHERPAV{2.1} signal samples  instead of the (nominal)
\POWHEGBOX+\PYTHIAV{8}. The effect of this uncertainty on $A_{W,Z}$
is estimated to be $\pm$0.9\%.

The systematic uncertainty in the multijet background estimation can be divided
into several components: the normalisation uncertainty from the $\chi^2$ fit,
the uncertainty in the modelling of electroweak processses by
simulated samples in  the fitted region, uncertainty from fit bias due to binning choice,
and uncertainty from template shape. The scale normalisation uncertainty from the $\chi^2$ fit is
approximately $\pm13$\%  for the $W\rightarrow e\nu$  channel. This
uncertainty is neglected in the $W \rightarrow \mu \nu$ channel where the
template bias is dominant.    The mismodelling uncertainty is estimated by
comparison  of the fit results for $ \ell^+$ and $\ell^-$, and for the combined $\ell^{\pm}$
candidates. 
The central value used is $0.5 N^{\pm}$ with the uncertainties $N^+ -
0.5N^{\pm}$ and $N^- - 0.5 N^{\pm}$, where $N^+$ is the fitted number of
$\ell^+$ background events, $N^-$ is the fitted number of
$\ell^-$ and $N^\pm$ is the fitted total number of
$\ell^{\pm}$ background events. In the $W \rightarrow e \nu$ channel this leads to an
uncertainty of $\pm 28$\% in the multijet background. In the
$W \rightarrow \mu \nu$ channel the multijet template normalisation is derived
from the fit in the small-$m_{\text{T}}$ region, where electroweak
contributions are negligible and there are many data events, and this
source of systematic error is found to be negligible. The fit-bias uncertainty arising from  the
choice of bin width  is estimated by repeating the fit with different
binnings. This component is negligible in the $W \rightarrow \mu \nu$ case and
$\pm 15$\% in the  $W \rightarrow e \nu$  case.  The uncertainty due
to a potential bias from  template choice is estimated by employing
different template selections. For the $W \rightarrow  e \nu$ channel, different
inverted-isolation criteria were investigated. The overall differences are
considered negligible. For the $W \rightarrow \mu \nu$ channel, template
variations were estimated from fits that use $b\bar{b}$ + $c\bar{c}$
MC samples as the multijet templates, leading to an uncertainty of
$\pm 75\%$; this is the largest uncertainty in the multijet background in the  $W \rightarrow \mu \nu$ channel.

\begingroup
\renewcommand*{\arraystretch}{1.1}
\begin{table}
\caption{\label{tab:WZcorrModelABC} The correlation model for the grouped systematic uncertainties for the
measurements of $W$-boson and $Z$-boson production. The entries in different rows are uncorrelated
with each other. Entries in a row with the same letter are fully correlated. Entries in
a row with a starred letter are mostly correlated with the entries with the same letter (most
of the individual sources of uncertainties within a group are taken as correlated). Entries
with different letters in a row are either fully or mostly uncorrelated with each other.}
\begin{center}
\begin{tabular}{l|ccc|ccc}
 
\multirow{2}{*}{Source}  & \multicolumn{3}{c|}{Muon channel} & \multicolumn{3}{c}{Electron channel}  \\
\cline{2-7}
& $Z$ & $W^{+}$  & $W^{-}$  & $Z$ & $W^{+}$  & $W^{-}$ \\
 
\hline
Muon trigger                    &  A       &  A      &  A       &   --   &   --   &   --  \\
Muon reconstruction/ID          &  A   &  A  &  A   &   --   &   --   &   --  \\
Muon energy scale/resolution    &  A       &  A      &  A       &   --   &   --   &   --  \\
Muon isolation    &  A       &  A      &  A       &   --   &   --   &   --  \\
\hline
Electron trigger                  &  --   &  --   & --   &   A$^{*}$   &   A$^{*}$   &   A$^{*}$  \\
Electron reconstruction/ID        &  --   &  --   &  --   &   A   &   A   &   A  \\
Electron energy scale/resolution  &  --   &  --   &  --   &   A       &   A       &   A      \\
Electron isolation  &  --   &  --   &  --   &   A       &   A       &   A      \\
\hline
Recoil related                        &  --   &  A   &  A   &   --   &   A   &   A  \\
\hline
EW background                    &  A   &  B   &  B   &   A   &   B   &   B  \\
Top-quark background                    &  A   &  A   &  A   &   A   &   A   &   A  \\
Multijet background                     &  --   &   A  &  A   &   --   &   A   &   A  \\
\hline
PDF                               &  A   &  A   &  A   &   A   &   A   &  A  \\

\end{tabular}
\end{center}
\end{table}
\endgroup

Combining results and building ratios or asymmetries of results require a model for the
correlations of particular systematic uncertainties between different
measurements. Correlations arise mostly due
to the  fact that electrons, muons, and the recoil are reconstructed
identically in the different measurements. Further correlations occur due to
similarities in the analysis methodology such as the methods of
signal and background estimation.
 
The systematic uncertainties from the electroweak background
estimations are treated as uncorrelated between the $W$-boson and $Z$-boson measurements,  and
fully correlated among different flavour decay channels of the $W$
and $Z$ boson.  The top-quark background is treated as fully correlated across
all $W$-boson and $Z$-boson  decay channels.  The multijet background and recoil-related systematic uncertainties are also treated as fully correlated between all four $W$-boson decay channels despite there being an expected uncorrelated component, since the statistical uncertainty  is dominant in this case.
 
The systematic uncertainties due to the choice of PDF are treated as
fully correlated between all $W$-boson and $Z$-boson channels.  The uncertainties in electron and muon selection, reconstruction and efficiency are treated as fully correlated between all $W$-boson and $Z$-boson channels.

A simplified form of the correlation model with the grouped list of the sources of systematic errors is presented in Table~\ref{tab:WZcorrModelABC}.
% End of text imported from the .//tex/Systematics.tex input file
 
% The next lines are included from the .//tex/Results.tex input file
\section{Results}
\label{sec:results}
 
\begin{figure}
\centering
\subfloat[]{\includegraphics[width=0.4 \textwidth]{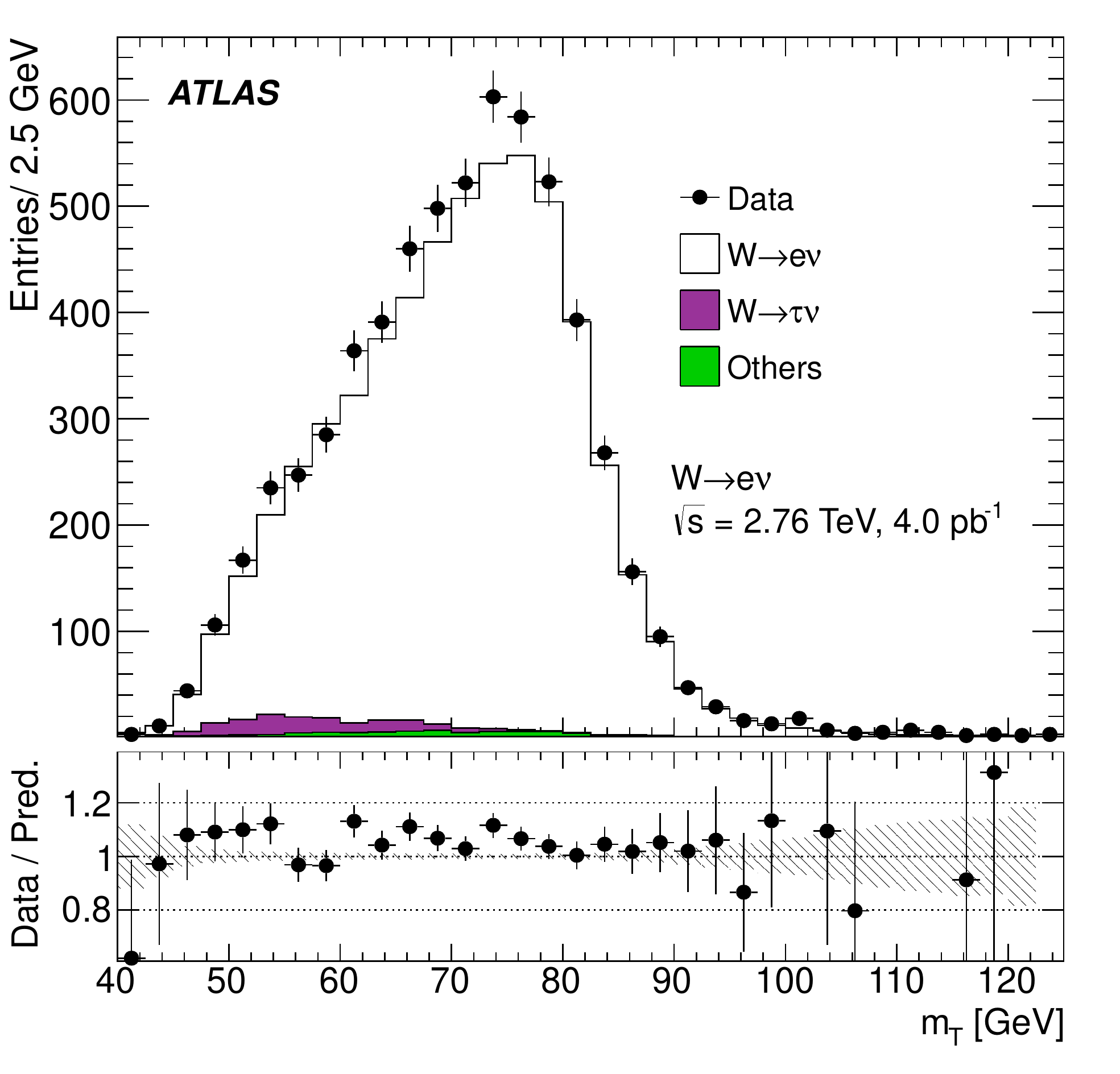}}
\subfloat[]{\includegraphics[width=0.4 \textwidth]{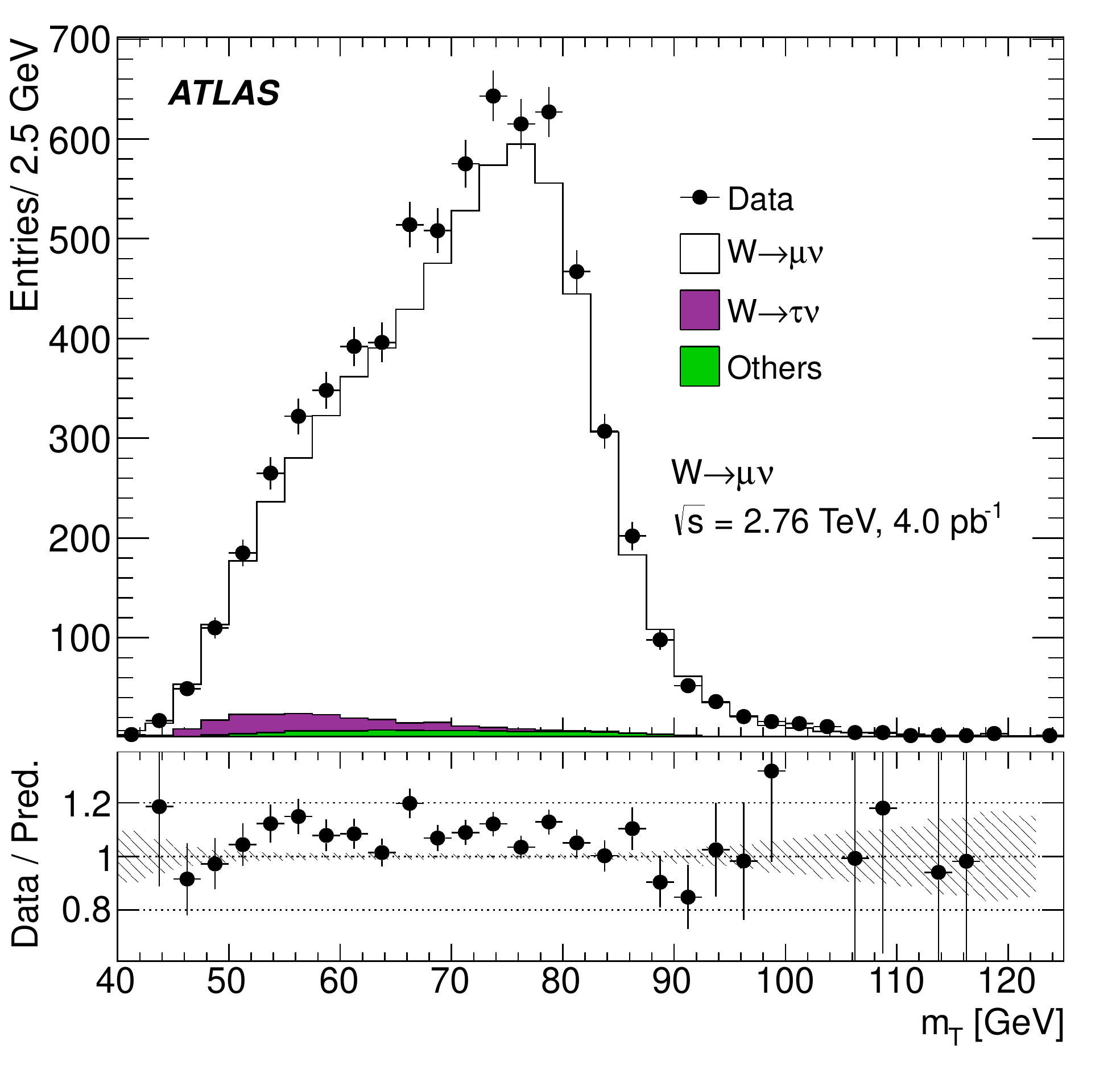}}
\caption{The distribution of $m_{\text{T}}$ for $W \rightarrow
\ell \nu$ candidate events. The expected signal, normalised to the
NNLO theoretical predictions, is shown as an unfilled histogram on
top of the stacked background predictions. Backgrounds that do not
originate from $W$ production are grouped together into the `Others' histogram.  Systematic uncertainties for the signal and background distributions are combined in the shaded band. Systematic uncertainties from the measurement of the integrated luminosity are not included. The lower panel shows the ratio of the data to the prediction.\label{fig:mtw}}
\end{figure}

\begin{figure}
\centering
\subfloat[]{\includegraphics[width=0.4 \textwidth]{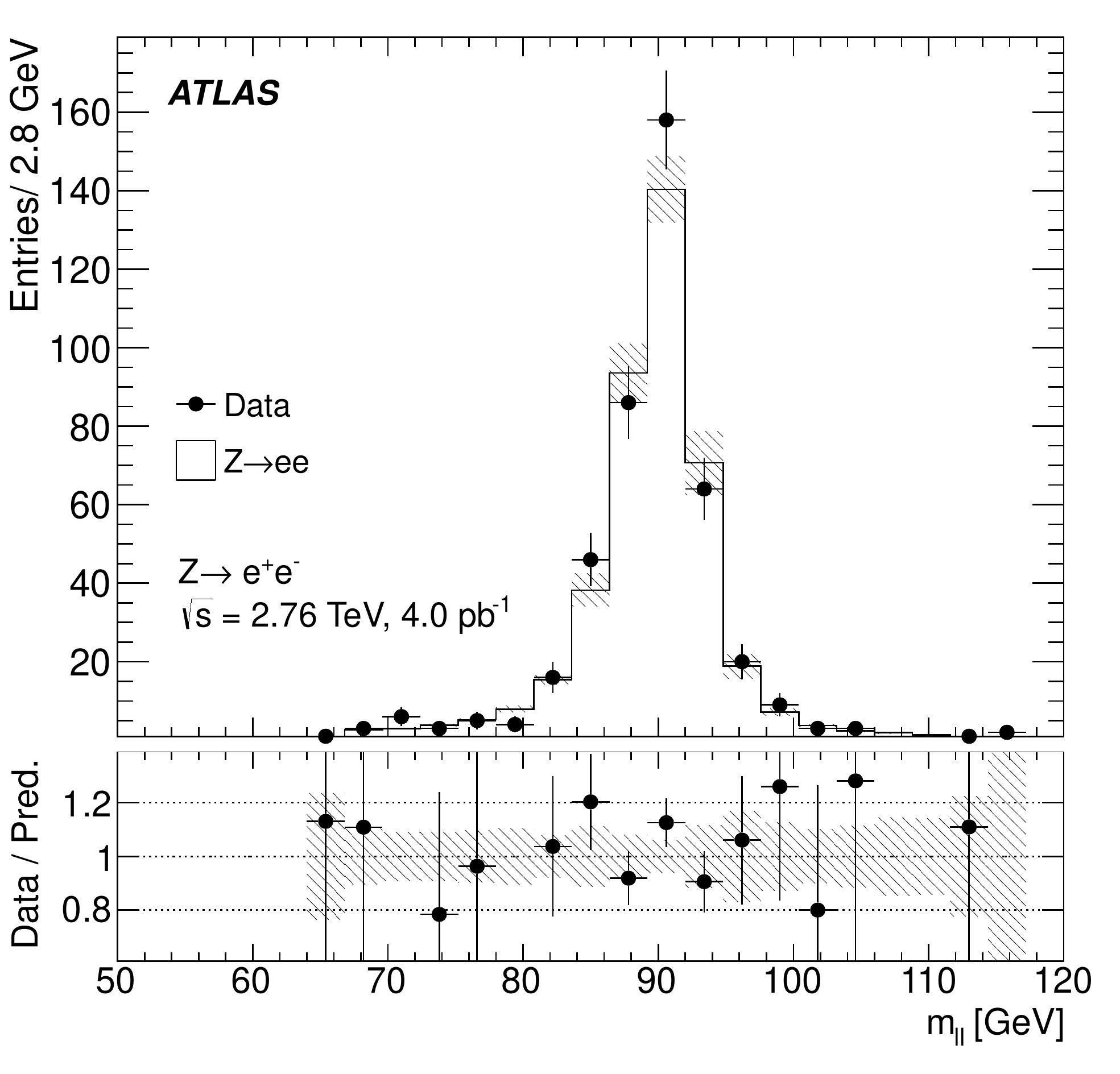}}
\subfloat[]{\includegraphics[width=0.4 \textwidth]{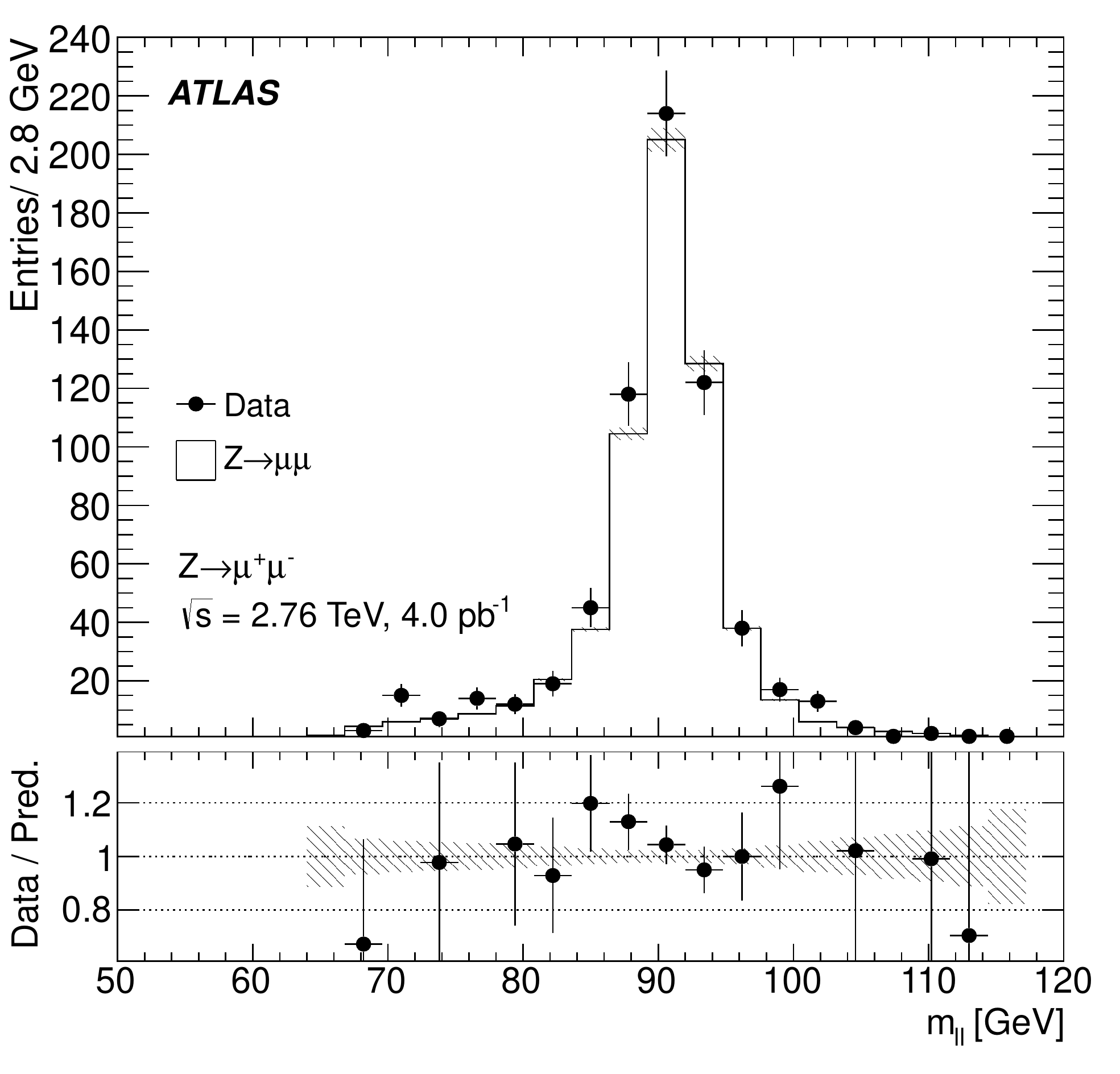}}
\caption{The distribution of $m_{\ell \ell}$ for $Z \rightarrow
\ell^+ \ell^-$ candidate events. The expected signal, normalised to the
NNLO theoretical predictions, is shown as an unfilled
histogram. Systematic uncertainties for the signal and background
distributions are combined in the shaded band. Systematic
uncertainties from the measurement of the integrated luminosity are
not included. The background distributions are neglected here, but
would not be visible if included. The lower panel shows the ratio of the data to the prediction.\label{fig:mll}}
\end{figure}

The numbers of events passing the event selections described in
Section~\ref{sec:selection} are presented in Table~\ref{tab:events}, together
with the estimated background contributions described in
Section~\ref{sec:backgrounds}. The distribution of $m_{\text{T}}$ for
$W \rightarrow \ell \nu$ candidate events is shown in Figure~\ref{fig:mtw},
compared with the expected distribution for signal plus backgrounds,
where the signal is normalised to the NNLO QCD prediction.
Similarly, Figure~\ref{fig:mll} shows the  distribution of $m_{\ell \ell}$ for
$Z \rightarrow \ell^+ \ell^-$ candidate events compared with the expectations for
signal. In this case, the background contributions are not shown,
because they would not be visible in the figure if included.

\begin{table}[!h]
\caption{The numbers of observed candidate events
with the estimated numbers of selected electroweak (EW) plus top, and multijet background
events, together with their total uncertainty.   In addition, the
number of  background-subtracted signal events  is shown  with the first  uncertainty
given being statistical and the second uncertainty being the
total systematic uncertainty, obtained by summing in quadrature
the EW+top and multijet uncertainties. Uncertainties shown as
$\pm 0.0 $ have a magnitude less than $0.05.$ \label{tab:events} }
\label{tab:BkgWlnu}
\begin{center}
\begin{tabular}{l | c | c | c | c  }
Measurement & Observed & Background & Background & Background-subtracted \\
Channel	 & candidates & (EW + top) & (Multijet) & data $N_{W}^{\text{sig}}$ \\
\hline
$W^+ \rightarrow e^{+} \nu$ & \ntotWplusenu & \nEWttbarbkgWplusenu~\hspace{0.1cm}  & \nQCDWplusenu & \ntotsignalWplusenu \\
$W^- \rightarrow e^{-} \bar{\nu}$ & \ntotWminenu & \nEWttbarbkgWminenu & \nQCDWminenu & \ntotsignalWminenu \\
$W^+ \rightarrow \mu^{+} \nu$ & \ntotWplusmunu &
\nEWttbarbkgWplusmunu~\hspace{0.1cm}
& \nQCDWplusmunu &
\ntotsignalWplusmunu~\hspace{0.1cm}  \\
$W^- \rightarrow \mu^{-} \bar{\nu}$ & \ntotWminmunu &\nEWttbarbkgWminmunu~\hspace{0.1cm} & \nQCDWminmunu &  \ntotsignalWminmunu~\hspace{0.1cm}  \\
\hline
$Z \rightarrow e^+ e^-$ & \ntotZee & \nEWttbarbkgZee  & -- &\ntotsignalZee \\
$Z \rightarrow \mu^+ \mu^-$ & \ntotZmumu &\nEWttbarbkgZmumu  & -- & \ntotsignalZmumu \\
\end{tabular}
\end{center}
\end{table}

The measured fiducial ($\sigma^{\text{fid}}$) and total ($\sigma^{\text{tot}}$)
cross-sections in the electron and muon channels are presented separately in
Table~\ref{tab:Wcs}. For these measurements, the dominant contribution
to the systematic uncertainty arises from the luminosity determination.
 
\begin{table}[!tb]
\caption{Results of the  fiducial  and total cross-sections measurements of the
$W^{+}$-boson, $W^{-}$-boson, and $Z$-boson production cross-sections in the electron and muon channels. The cross-sections are
shown with their statistical, systematic and luminosity uncertainties (and
extrapolation uncertainty for total cross-section).}
\label{tab:Wcs}
\begin{center}
\begin{tabular}{ l | c| c }
& Value $\pm$ stat. $\pm$ syst. $\pm$ lumi. ($\pm$ extr.)& Value $\pm$ stat. $\pm$ syst.
$\pm$ lumi. ($\pm$ extr.) \\
\hline
& $W^{+}\to e\nu$ & $W^{+}\to \mu\nu$ \\
 
\hline
$\sigma^{\text{fid}}_{W^+}$ [pb]  & \sigfidWplusenunolabel \hspace{0.9cm}\hspace{0.15cm}&
\sigfidWplusmununolabel \hspace{0.9cm}\hspace{0.15cm}\\
$\sigma^{\text{tot}}_{W^+}$ [pb] & \hspace{0.15cm}\sigtotWplusenunolabel &  \hspace{0.25cm}\sigtotWplusmununolabel\hspace{0.05cm} \\
\hline
& $W^{-}\to e\nu$ & $W^{-}\to \mu\nu$\\
\hline
$\sigma^{\text{fid}}_{W^-}$ [pb] & \hspace{0.15cm} \sigfidWminenunolabel \hspace{1.0cm} & \hspace{0.15cm} \sigfidWminmununolabel \hspace{0.8cm}\hspace{0.15cm} \\
$\sigma^{\text{tot}}_{W^-}$ [pb]  & \hspace{0.10cm}\sigtotWminenunolabel & \hspace{0.05cm}\sigtotWminmununolabel \\
\hline
& $Z \to ee$ & $ Z \to \mu\mu$ \\
\hline
$\sigma^{\text{fid}}_{Z}$  [pb] &\sigfidZeenolabel \hspace{1.15cm} &
\sigfidZmumunolabel
\hspace{1.20cm}
\\
$\sigma^{\text{tot}}_{Z}$  [pb]  & \sigtotZeenolabel  &  \sigtotZmumunolabel \\
\end{tabular}
\end{center}
\end{table}

The results obtained from the electron and muon final states are
consistent. The fiducial measurements from electron and muon final
states are combined following the procedure described in
Ref.~\cite{Glazov:2005rn} and the result is extrapolated to the full
phase space to obtain the total cross-section. The total  $W$-boson cross-section is calculated by summing the separate $W^+$ and $W^-$ cross-sections. The results are shown in Table~\ref{tab:CsComb}.

\begin{table}[!tbp]
\caption{Combined  fiducial  and total cross-section measurements for
$W^{+}$-boson, $W^{-}$-boson and $Z$-boson production. The cross-sections are
shown with their statistical, systematic and luminosity uncertainties (and
extrapolation uncertainty for total cross-section).}
\label{tab:CsComb}
\begin{center}
\begin{tabular}{ l | c | c }
& Value $\pm$ stat. $\pm$ syst. $\pm$ lumi. ($\pm$ extr.)& Value $\pm$ stat. $\pm$ syst.
$\pm$ lumi. ($\pm$ extr.) \\
\hline
& $W^{+}\to \ell \nu$ & $W^{-}\to \ell \nu$ \\
 
\hline
$\sigma^{\text{fid}}_{W}$ [pb]  & $\valfidWp \pm \statfidWp \pm \sysfidWp \pm \lumifidWp$\hspace{0.9cm}\hspace{0.15cm} &
\hspace{0.35cm}$\valfidWm \pm \statfidWm \pm \sysfidWm \pm \lumifidWm$\hspace{0.2cm}\hspace{1.0cm}  \\
$\sigma^{\text{tot}}_{W}$ [pb] & $\valAfullWp \pm \statAfullWp \pm \sysAfullWp \pm
\lumiAfullWp \hspace{0.15cm} (\pm \extrAfullWp$) & $\valAfullWm \pm \statAfullWm \pm \sysAfullWm \pm
\lumiAfullWm \hspace{0.15cm} (\pm \extrAfullWm$) \\
\hline
& \multicolumn{2}{c}{$W \to \ell \nu$} \\
\hline
$\sigma^{\text{fid}}_{W}$ [pb] & \multicolumn{2}{c}{$\valfidWtot \pm \statfidWtot \pm
\sysfidWtot \pm \hspace{0.2cm}\lumifidWtot$}\hspace{0.8cm}\hspace{0.15cm}  \\
$\sigma^{\text{tot}}_{W}$ [pb]  & \multicolumn{2}{c}{$\valAfullWtot \pm \statAfullWtot \pm
\sysAfullWtot \pm \lumiAfullWtot \hspace{0.15cm} (\pm \extrAfullWtot$)} \\
\hline
& \multicolumn{2}{c}{$Z \to \ell \ell$} \\
\hline
$\sigma^{\text{fid}}_{Z}$ [pb] & \multicolumn{2}{c}{\hspace{0.05cm}$\valfidZ \pm \statfidZ \pm \sysfidZ \pm
\hspace{0.20cm} \lumifidZ$}\hspace{1.0cm}\hspace{0.15cm} \\
$\sigma^{\text{tot}}_{Z}$ [pb]  & \multicolumn{2}{c}{$\valAfullZ \pm \statAfullZ \pm
\sysAfullZ \pm \lumiAfullZ \hspace{0.15cm} (\pm \extrAfullZ$)} \\
\end{tabular}
\end{center}
\end{table}

Theoretical predictions of the fiducial and total cross-sections are computed
for comparison with the measured cross-sections using
\textsc{Dynnlo} 1.5~\cite{Catani:2007vq}
which provides calculations at NNLO in the strong-coupling
constant, $\mathcal{O}(\alpha_\mathrm{s}^2)$, including the boson decays
into leptons ($\ell^+\nu, \ell^- \bar{\nu}$ or $\ell^+ \ell ^-$) with
full spin correlations, finite width and interference
effects. These calculations
allow  kinematic requirements to be implemented for direct
comparison with experimental data.  The procedure used follows that used for
the previous ATLAS measurement at
$\sqrt{s}=7$~\TeV{}~\cite{Aaboud:2016btc}.
 
Corrections for NLO EW effects are calculated with \textsc{Fewz} 3.1~\cite{Melnikov:2006kv,Gavin:2010az,Gavin:2012sy,Li:2012wna}, for the $Z$ bosons and with
\textsc{Sanc}~\cite{Bardin:2012jk,Arbuzov:2012dx} for the $W$ bosons.
The calculation was done in the $G_{\mu}$ EW scheme~\cite{Hollik:1988ii}. The
following input parameters are taken from the Particle Data Group's
Review of Particle Properties 2014 edition~\cite{PDG14}: the Fermi constant, the masses and widths of $W$ and $Z$ bosons
as well as the elements of the CKM matrix.
The cross-sections for vector bosons decaying into these leptonic
final states are calculated such that they match the definition of the measured cross-sections in the data.
Thus, from complete NLO EW corrections, the following components are included:
virtual QED and weak corrections, real initial-state radiation (ISR),
and interference between ISR and real final-state radiation
(FSR)~\cite{Dittmaier:2009cr}. The calculated effect of these
corrections on the cross-sections is $(-0.26 \pm 0.02 )$\% for
$\sigma^{\text{fid}}_{W^+}$,   $(-0.21 \pm 0.03) $\% for
$\sigma^{\text{fid}}_{W^-}$, and  $(-0.25 \pm 0.12) $\% for
$\sigma^{\text{fid}}_{Z}$.
\textsc{Dynnlo} is used for the central values of the predictions while
$\textsc{Fewz}$ is used for the PDF variations
and all other systematic variations such as QCD scale and $\alpha_\mathrm{s}$. The predictions are calculated using the
CT14nnlo~\cite{Dulat:2015mca}, NNPDF3.1~\cite{Ball:2017nwa}, MMHT14nnlo68cl~\cite{Harland-Lang:2014zoa}, ABMP16~\cite{Alekhin:2017kpj}, HERAPDF2.0~\cite{Abramowicz:2015mha}, and
ATLAS-epWZ12nnlo PDF sets. The dynamic scale, $m_{\ell \ell}$, and
fixed scale, $m_W$, are used as the
nominal renormalisation, $\mu_\mathrm{R}$, and factorisation, $\mu_\mathrm{F}$, scales
for $Z$ and $W$ predictions, respectively.
 
Theoretical uncertainties in the predictions are also
derived from the following sources:
\begin{description}
\item PDF:\,these uncertainties are evaluated from the variations of
the NNLO PDFs according to the recommended procedure for
each PDF set. A table with all PDF uncertainties and their central
values is shown in Appendix~\ref{app:theory}; the PDF uncertainty
from CT14nnlo was rescaled from 90\%~CL to 68\%~CL.
\item\,Scales:\,the scale uncertainties are defined by the envelope of the variations in which the scales are changed by factors of two subject to the constraint $0.5 \le \mu_\mathrm{R}/\mu_\mathrm{F} \le 2$.
\item\,$\alpha_\mathrm{s}$:~the uncertainty due to $\alpha_\mathrm{s}$ was estimated
by varying the value of $\alpha_\mathrm{s}$ used in the CT14nnlo PDF set by $\pm 0.001$, corresponding to a 68\% CL variation.
\end{description}
The statistical uncertainties in these theoretical predictions are negligible.

\begin{table}[!h]
\caption{The predictions, using the CT14nnlo PDF set, for the
cross-sections measured. The calculations are
performed using \textsc{Dynnlo} 1.5 and \textsc{Fewz} 3.1
as described in the text. The errors represent the PDF and scale uncertainties.\label{tab:ctpred} }
\begin{center}
\begin{tabular}{  c | c   }
Quantity & Predicted cross-section [pb]\\
\hline
\rule[-2mm]{0mm}{6mm}
$\sigma_{W^+}^{\text{fid}}$&$1379 \hspace{0.25cm}\ ^{+39 \hspace{0.2cm}}_{-40 \hspace{0.2cm}}\ ^{+\hspace{0.2cm} 6 \hspace{0.1cm}}_{-\hspace{0.2cm} 6 \hspace{0.1cm}}  $~\hspace{0.1cm}\\
\rule[-2mm]{0mm}{6mm}          $\sigma_{W^-}^{\text{fid}}$&
\hspace{0.3cm}$757.3\ ^{+20.5}_{-24.5}\ ^{+\hspace{0.15cm} 3.1}_{-\hspace{0.15cm} 3.1} $\hspace{0.15cm} \\
\rule[-3mm]{0mm}{7mm}
$\sigma_{Z}^{\text{fid}}$&\hspace{0.3cm}$196.0\ ^{+\hspace{0.15cm}5.0}_{-\hspace{0.15cm}5.8}\ ^{+\hspace{0.15cm} 1.1}_{-\hspace{0.15cm} 1.3}$\hspace{0.15cm} \\
\hline
\rule[-2mm]{0mm}{6mm}
$\sigma_{W^+}^{\text{tot}}$&$2115  \hspace{0.25cm}\ ^{+57 \hspace{0.25cm}}_{-60} \ ^{+\hspace{0.15cm} 9 \hspace{0.2cm}}_{-11}$ \\
\rule[-2mm]{0mm}{6mm}
$\sigma_{W^-}^{\text{tot}}$&$1266 \hspace{0.25cm}\ ^{+32 \hspace{0.25cm}}_{-38} \ ^{+\hspace{0.15cm} 5 \hspace{0.2cm}}_{-\hspace{0.15cm} 6 \hspace{0.2cm}}$\\
\rule[-3mm]{0mm}{6mm}
$\sigma_{Z}^{\text{tot}}$&\hspace{0.2cm}$304.1\hspace{0.05cm}^{+\hspace{0.15cm}7.3}_{-\hspace{0.15cm} 8.2} \ ^{+\hspace{0.15cm}1.1}_{-\hspace{0.15cm}1.4}$\hspace{0.2cm}\\
\end{tabular}
\end{center}
\end{table}
 
\begin{figure}[!tbp]
\centering
\subfloat[]{\includegraphics[width=0.8 \textwidth]{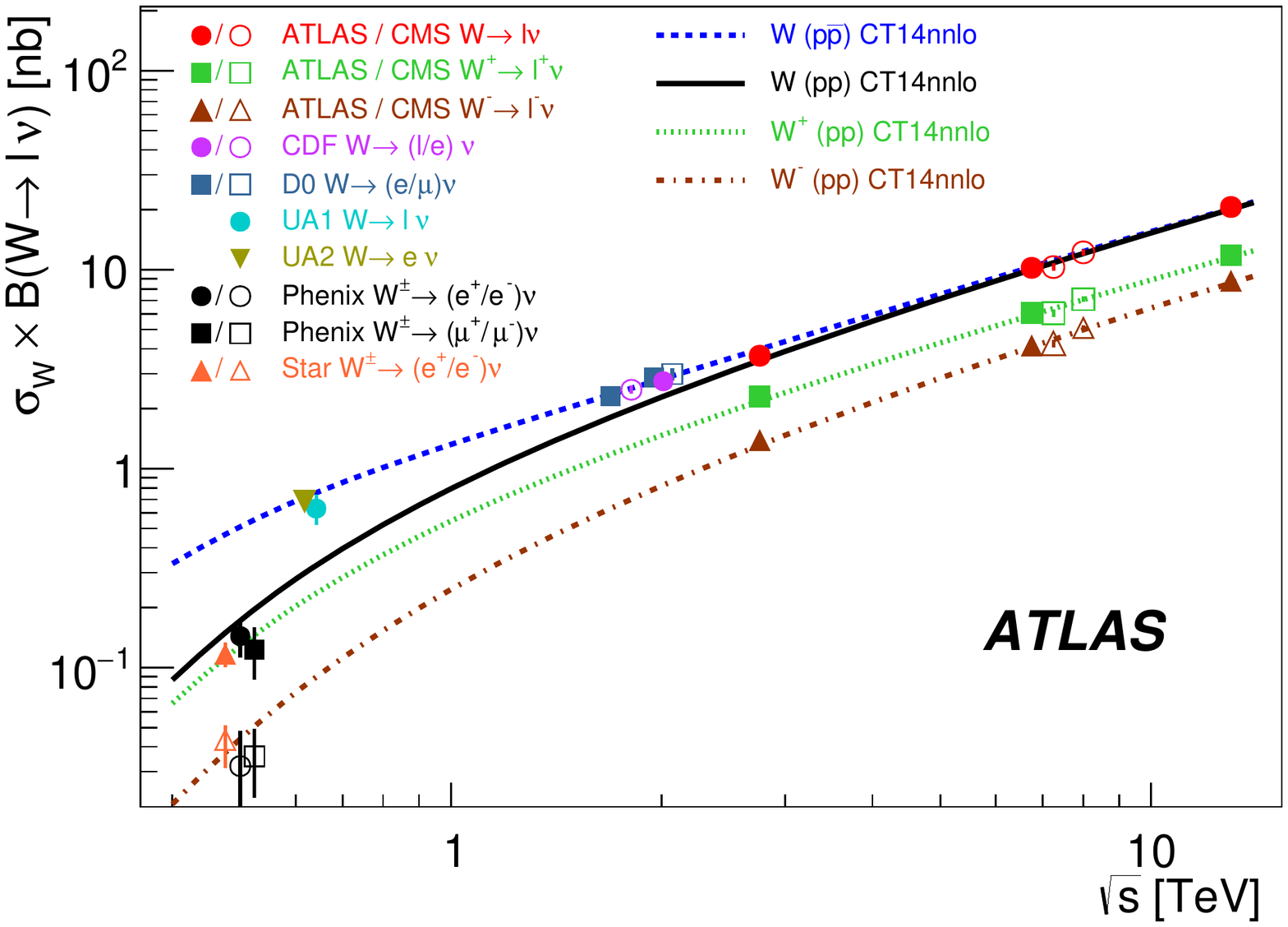}} \\
\subfloat[]{\includegraphics[width=0.8 \textwidth]{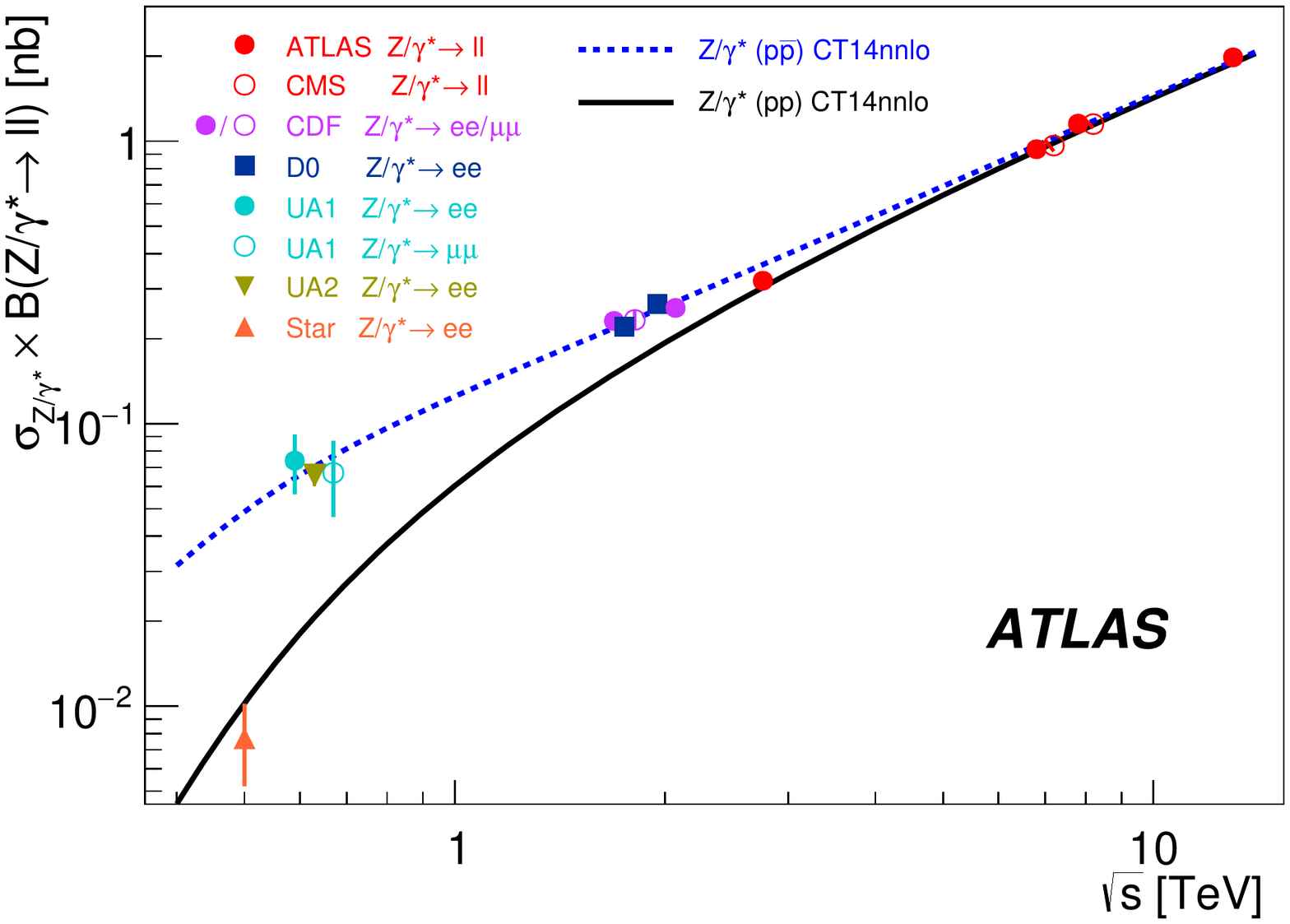}}
\caption{The measured values of (a) $\sigma_{W} \times B (W \rightarrow \ell \nu)$
for $W^+$  bosons, $W^-$  bosons and their sum and (b) $\sigma_{Z/\gamma^*} \times
B( Z/\gamma^*    \rightarrow \ell \ell)$ for proton--proton and proton--antiproton
collisions as a function of $\sqrt{s}$. Data points at the same
$\sqrt{s}$ are staggered to improve readibility. All data points are
shown together with their total uncertainty. The theoretical calculations
are performed at NNLO in QCD using \textsc{Dynnlo} 1.5 and \textsc{Fewz} 3.1
as described in the text. The theoretical uncertainties are not shown.\label{fig:xsecres}}
\end{figure}

\begin{figure}[!tbp]
\begin{minipage}[h]{0.5\linewidth}
\center{\includegraphics[width=1\textwidth]{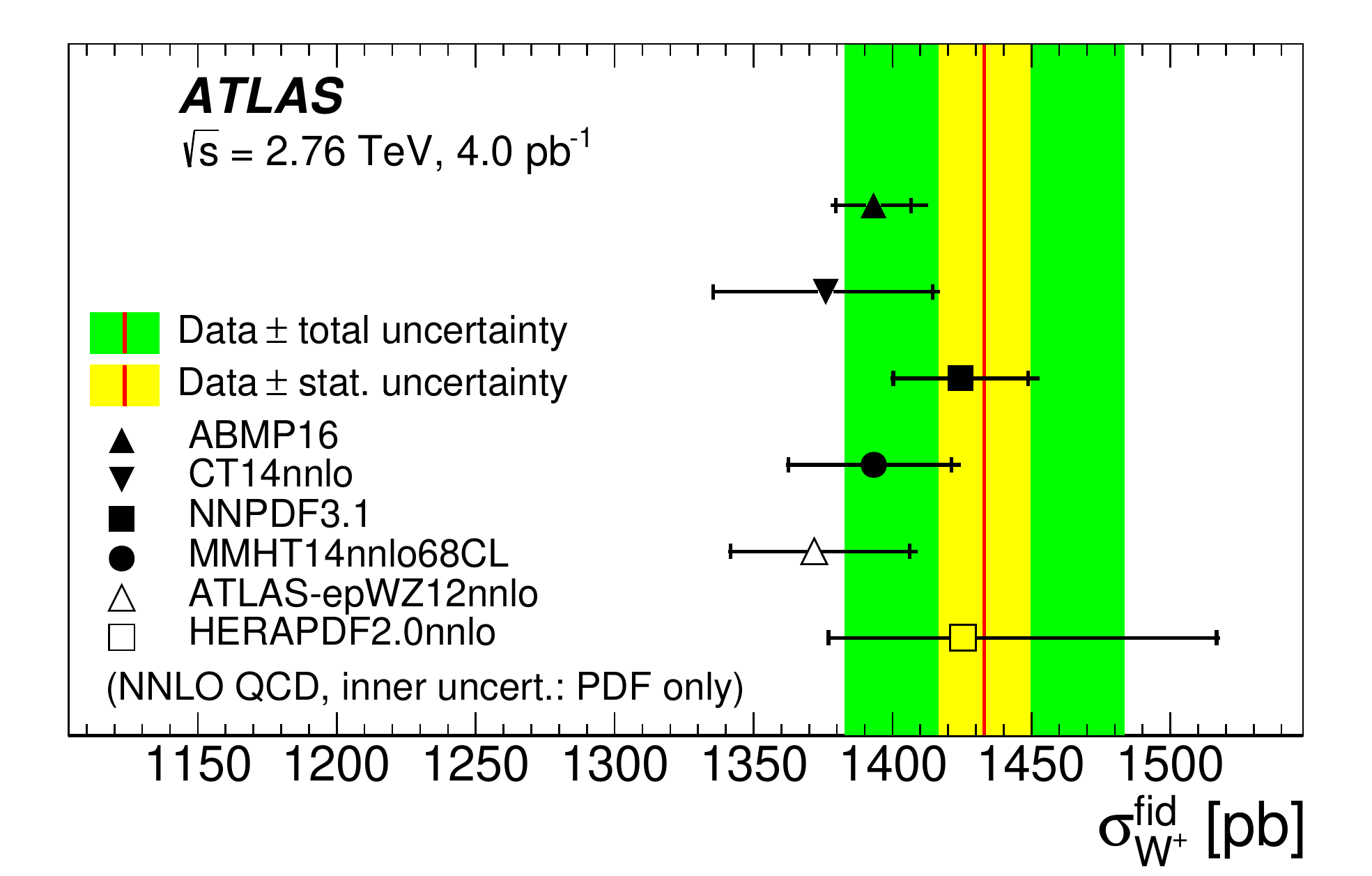} \\ (a)}
\end{minipage}
\hfill
\begin{minipage}[h]{0.5\linewidth}
\center{\includegraphics[width=1\textwidth]{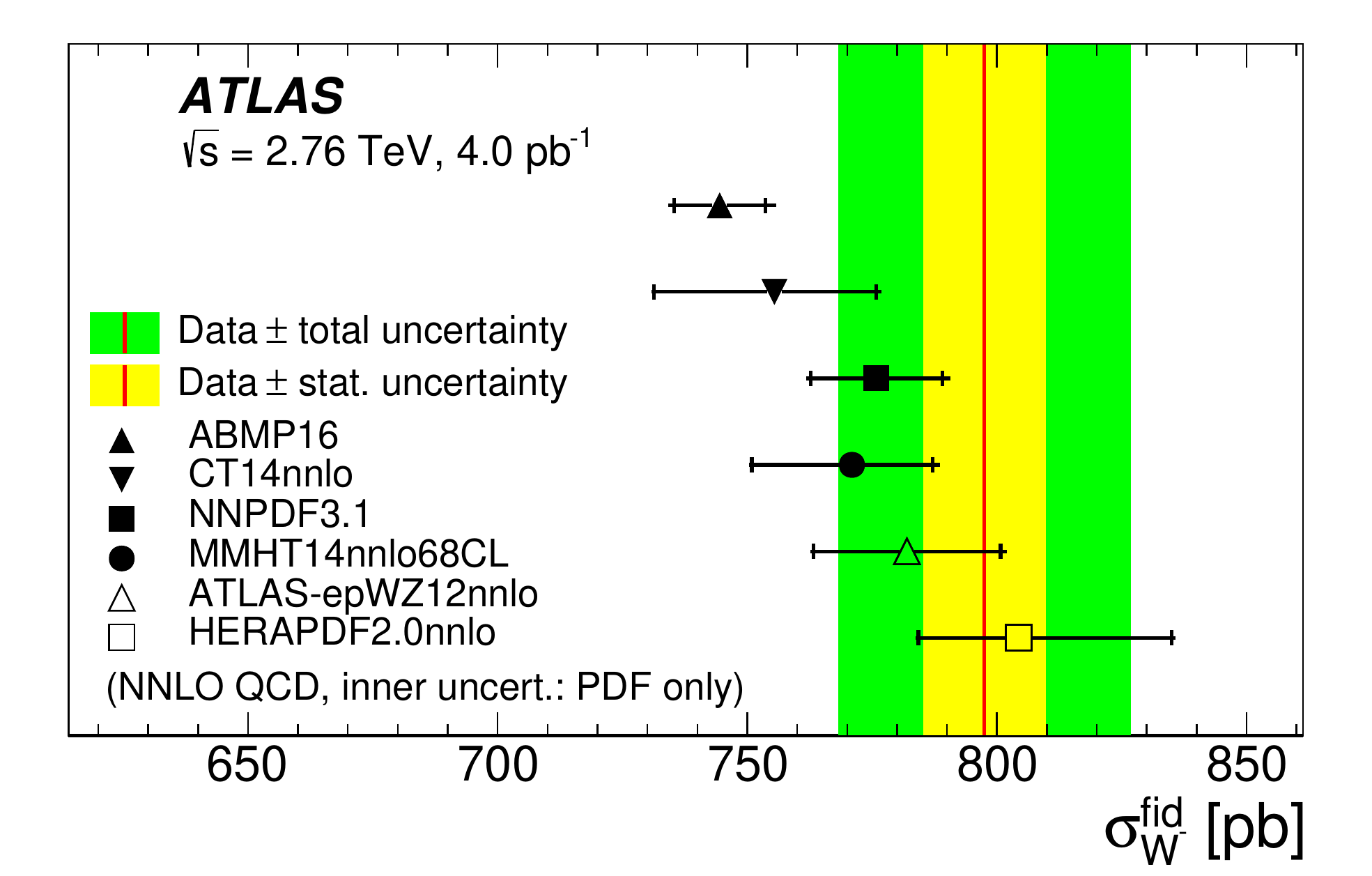} \\ (b)}
\end{minipage}
\caption{ NNLO predictions for the fiducial cross-section  (a) $\sigma^{\text{fid}}_{W^{+}}$ and (b) $\sigma^{\text{fid}}_{W^{-}}$ for the six
PDFs CT14nnlo, MMHT2014, NNPDF3.1, ATLASepWZ12, ABMP16 and, HERApdf2.0 compared with the measured fiducial cross-section as given
in Table~\ref{tab:CsComb}. The inner shaded band represents the statistical uncertainty only, the outer band corresponds to the
experimental uncertainty (including the luminosity uncertainty). The theory predictions are given with the corresponding
PDF (total) uncertainty shown by inner (outer) error bar.}
\label{fig:WpWmXsec}
\end{figure}

\begin{figure}[!tbp]
\begin{minipage}[h]{0.5\linewidth}
\center{\includegraphics[width=1\textwidth]{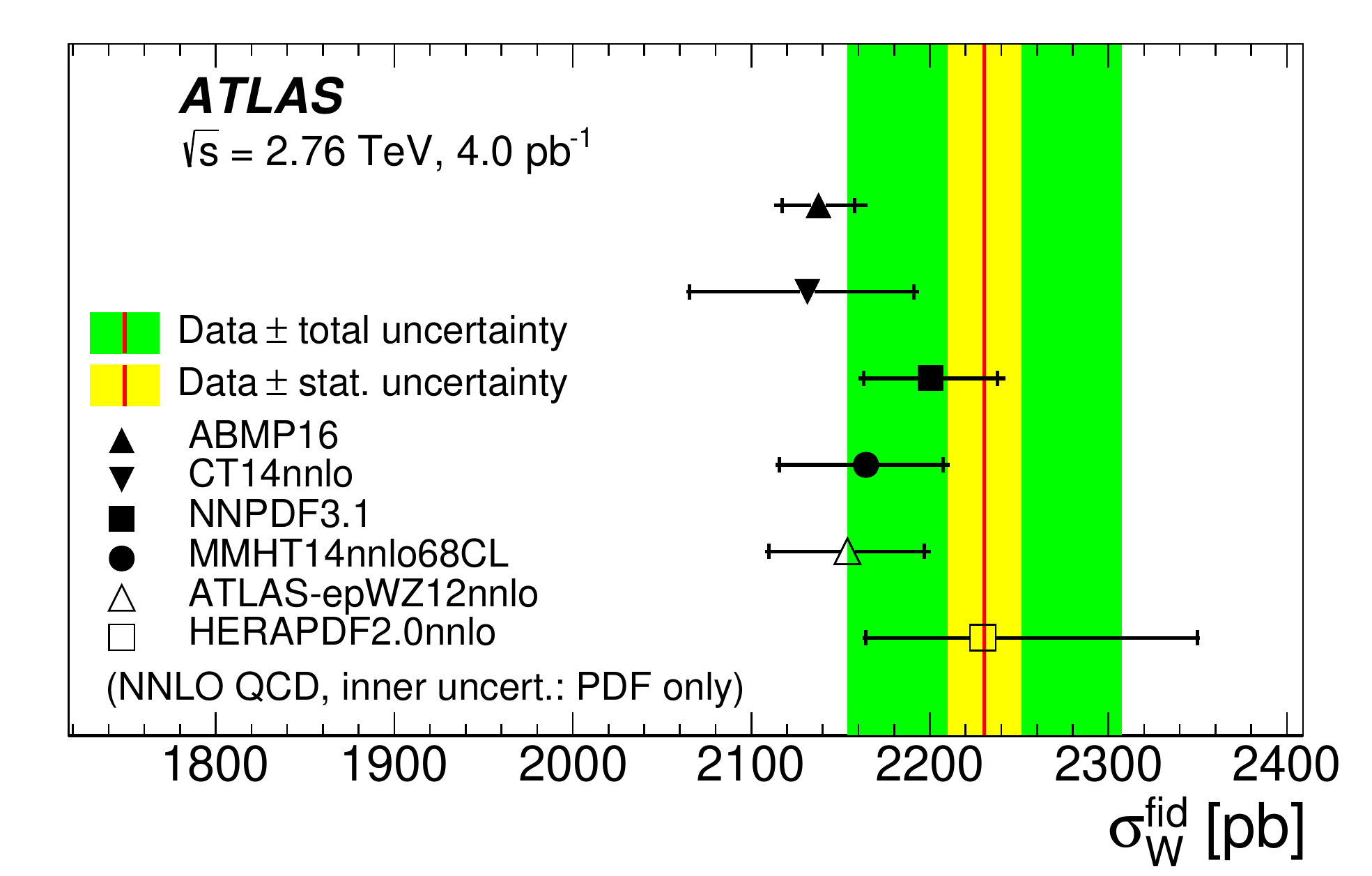} \\ (a)}
\end{minipage}
\hfill
\begin{minipage}[h]{0.5\linewidth}
\center{\includegraphics[width=1\textwidth]{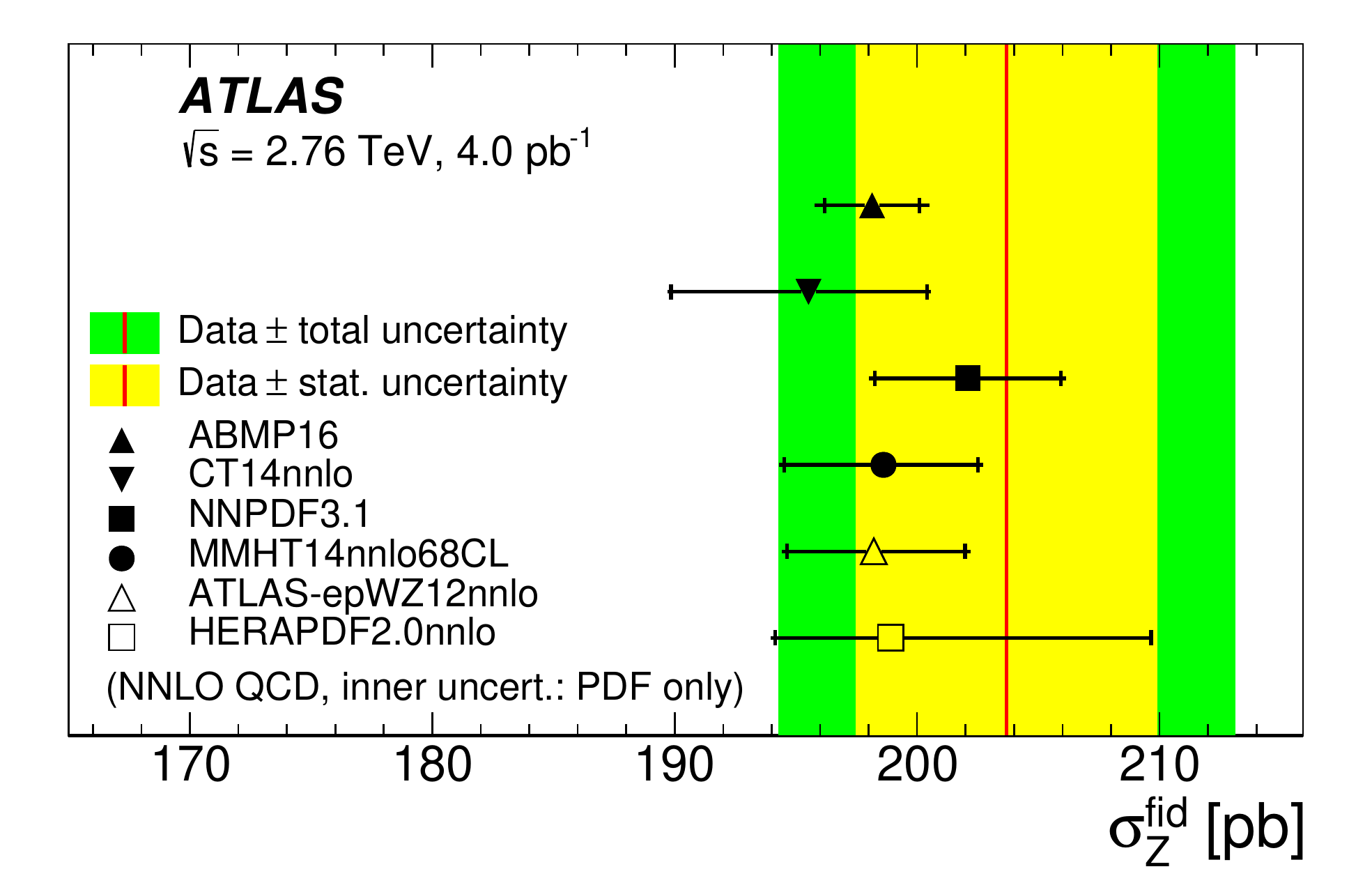} \\ (b)}
\end{minipage}
\caption{ NNLO predictions for the fiducial cross-sections  (a) $\sigma^{\text{fid}}_{W}$ and  (b) $\sigma^{\text{fid}}_{Z}$ for the six
PDFs CT14nnlo, MMHT2014, NNPDF3.1, ATLASepWZ12, ABMP16 and HERApdf2.0 compared with the measured fiducial cross-section as given
in Table~\ref{tab:CsComb}. The inner shaded band represents the statistical uncertainty only, the outer band corresponds to the
experimental uncertainty (including the luminosity
uncertainty). The theory predictions are given with the corresponding
PDF (total) uncertainty shown by inner (outer) error bar.}
\label{fig:WZXsec}
\end{figure}

The numerical values of the predictions for the CT14nnlo PDF set are
presented in Table~\ref{tab:ctpred}. The predictions for the
acceptance factor $A_{W,Z}$  can
differ by a few percent from those derived from simulated signal
samples, this may be due to a poorer description of
production of
low $p_{\text{T}}$ $W$-bosons by the fixed-order calculations. The predictions are shown in comparison with the combined $W$-boson and $Z$-boson production
measurements, and with results from $pp$ and $p\bar{p}$ collisions at other
centre-of-mass energies in Figure~\ref{fig:xsecres}. A comparison of the
measurements with predictions from various different PDF sets is presented in
Figures~\ref{fig:WpWmXsec} and \ref{fig:WZXsec}. Overall there is good
agreement.

\begin{figure}
\centering
\subfloat[]{\includegraphics[width=0.5 \textwidth]{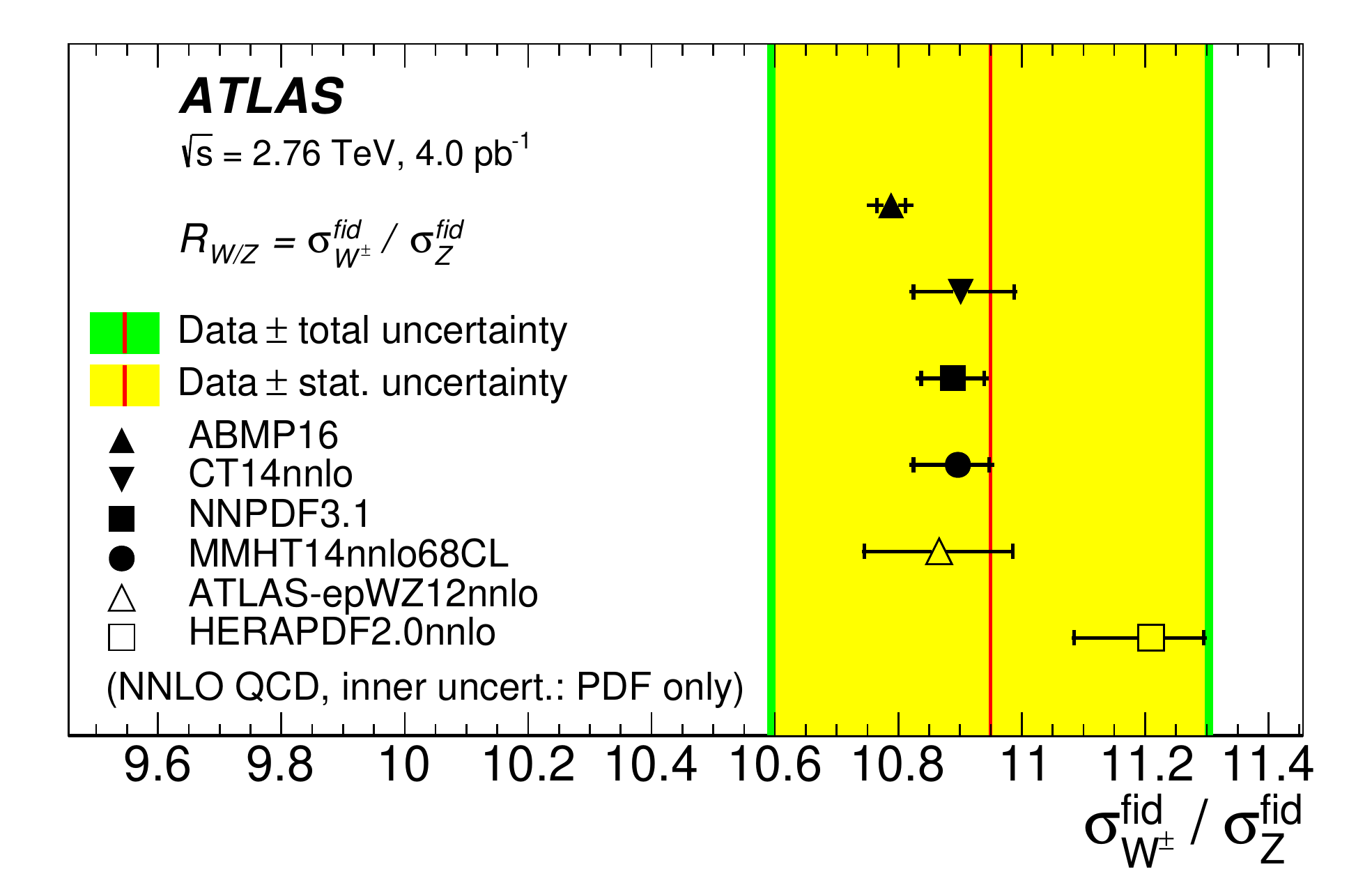}}
\subfloat[]{\includegraphics[width=0.5 \textwidth]{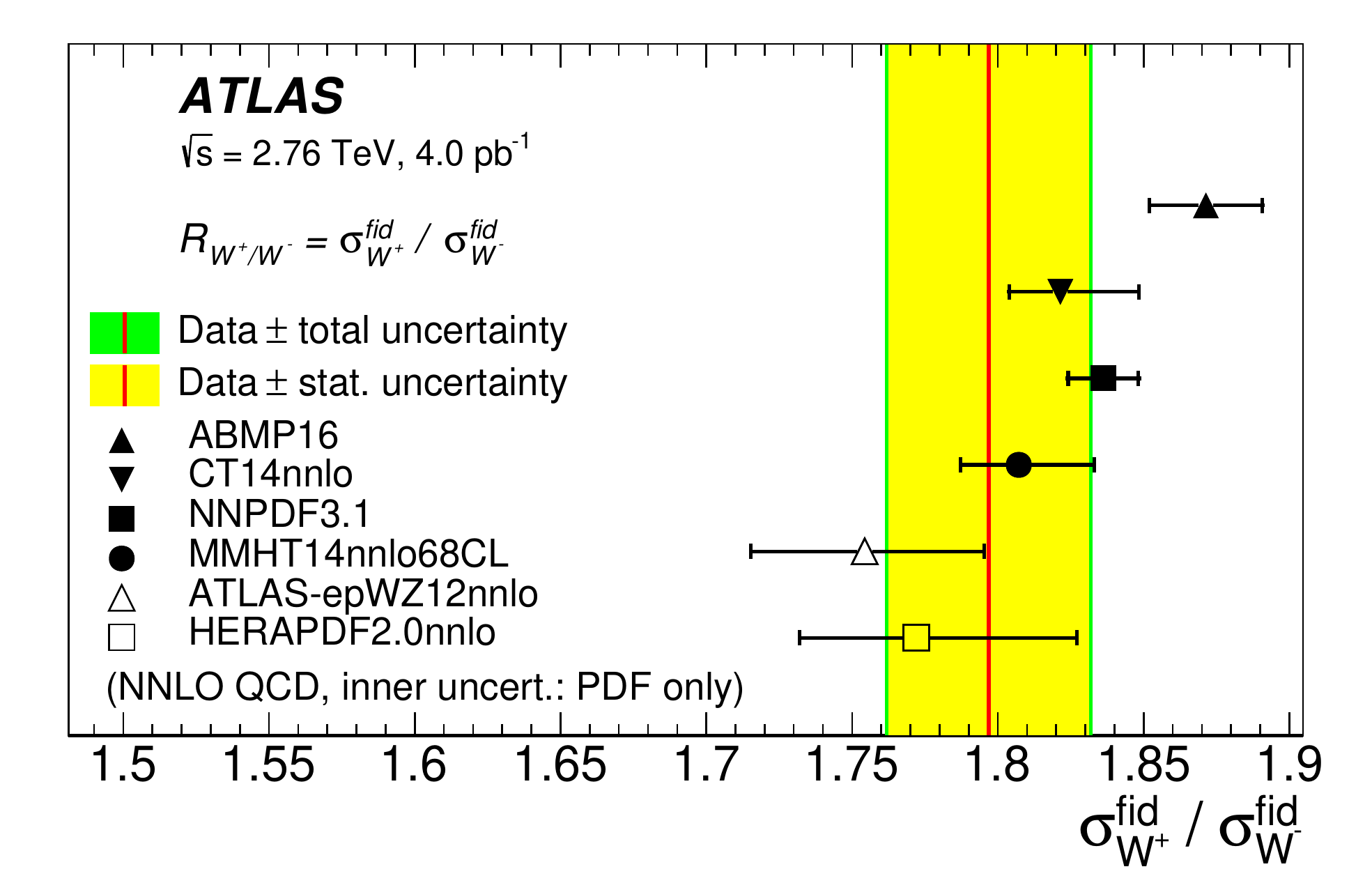}} \\
 
\caption{The measured ratio of fiducial cross-sections for (a)
$W$-boson production to $Z$-boson production, (b)  $W^+$-boson
production to $W^-$-boson production. The measurements are compared
with theoretical predictions  at NNLO in QCD based on  a selection of
different PDF sets. The inner shaded band corresponds to statistical
uncertainty while the outer band shows statistical and
systematic uncertainties added in quadrature. The theory predictions are given with the corresponding
PDF (total) uncertainty shown by inner (outer) error bar. \label{fig:xsecratios}}
\end{figure}

Taking ratios of measurements leads to results that have significantly reduced
systematic uncertainties due to full or partial cancellation of correlated systematic uncertainties, as
discussed  in  Section~\ref{sec:systematics}. The ratios of the fiducial cross-sections for  $W$-boson and $Z$-boson production are presented, together with the
ratio for $W^+$-boson
and $W^-$-boson production, in Figure~\ref{fig:xsecratios}. It can be seen that
the predictions from the different PDF sets are mostly in good agreement with the
measurements. There is a slight (less than two standard deviations) tension between the
data and the prediction using the ABMP16 PDF set. The measured values of the ratios are:

\begin{center}
$R_{W/Z} = \WZcombVal \pm \WZcombSt \, (\text{stat.})\, \pm \WZcombSy\, (\text{syst.})$; \\
$R_{W^{+}/W^{-}} = \RatCombCentWpoverWm \pm \RatCombStatWpoverWm\, (\text{stat.})\, \pm \RatCombSystWpoverWm\,  (\text{syst.})$. \\
\end{center}
 
The measurement of the ratio $R_{W^{+}/W^{-}}$is sensitive to the
$u_\mathrm{v}$ and
$d_\mathrm{v}$ valence quark distributions, while the ratio $R_{W/Z}$ can place
constraints on the strange quark distributions. A common alternative
way of presenting this information is in terms of the charge asymmetry, $A_\ell$, in $W$-boson production:
 
\begin{equation*}
A_{\ell} = \frac{ \sigma_{W^+}^{\text{fid}} - \sigma_{W^-}^{\text{fid}} }{ \sigma_{W^+}^{\text{fid}} + \sigma_{W^-}^{\text{fid}} }.
\end{equation*}
 
This observable also benefits  from the cancellation of systematic
uncertainties in the same way as the cross-section ratios. The
measured value is:
 
\begin{center}
 
$ A_{\ell} =  \WAsymcombVal \pm \WAsymcombSt (\text{stat.}) \pm \WAsymcombSy (\text{syst.})  $. \\
 
\end{center}

The ratio of measured cross-sections in the electron and muon decay
channels provides a test of lepton universality in $W$-boson
decays. The measured ratios are:
 
\begin{equation}
R_{W^+} = \frac{\sigma^{\text{fid}}_{W^+\rightarrow e^+
\nu}}{\sigma^{\text{fid}}_{W^+\rightarrow \mu^+ \nu}}
=0.985
\pm   0.023\
(\text{stat.})\
\pm 0.028\
(\text{syst.})  \nonumber
\end{equation}
 
\begin{equation}
R_{W^-} = \frac{\sigma^{\text{fid}}_{W^-\rightarrow e^-
\bar{\nu}}}{\sigma^{\text{fid}}_{W^-\rightarrow \mu^- \bar{\nu}}}
=0.988
\pm   0.030\
(\text{stat.})\
\pm 0.028\
(\text{syst.})  \nonumber
\end{equation}
 
\begin{equation}
R_{W} = \frac{\sigma^{\text{fid}}_{W\rightarrow e
\nu}}{\sigma^{\text{fid}}_{W\rightarrow \mu \nu }}
=0.986
\pm   0.018\
(\text{stat.})\
\pm 0.028\
(\text{syst.})  \nonumber
\end{equation}
 
\begin{equation}
R_{Z} = \frac{\sigma^{\text{fid}}_{Z\rightarrow e^+e^-}}{\sigma^{\text{fid}}_{Z\rightarrow \mu^+ \mu^- }}
=0.96
\pm   0.06\
(\text{stat.})\
\pm 0.05\
(\text{syst.})  \nonumber
\end{equation}
 
These results lie within one  standard deviation of
the Standard Model prediction and previous measurements by ATLAS.

% End of text imported from the .//tex/Results.tex input file
 
\FloatBarrier
% The next lines are included from the .//tex/Conclusion.tex input file
\section{Conclusion}
\label{sec:conclusion}

This paper presents measurements of the \wlnu and  \zll  production cross-sections
based on about
12~400 $W$-boson and 1100 $Z$-boson candidates, after subtracting
background events, reconstructed from
$\sqrt{s}=$ 2.76~\TeV{} proton--proton collision data recorded by the
ATLAS detector at
the LHC, corresponding to integrated luminosity of 4.0~pb$^{-1}$.
The total inclusive $W$-boson production cross-sections for the combined  electron and muon channels are
\begin{center}
$\sigma^{\text{tot}}_{W^+ \to \ell\nu} = \valAfullWp  \pm \statAfullWp\ (\text{stat.})\ \pm \sysAfullWp\ (\text{syst.}) \pm \lumiAfullWp\ (\text{lumi.})  \pm \extrAfullWp\  (\text{extr.}) \, \text{pb}$,\\
$\sigma^{\text{tot}}_{W^- \to \ell\nu} = \valAfullWm  \pm \statAfullWm\ (\text{stat.})\ \pm \sysAfullWm\ (\text{syst.}) \pm \lumiAfullWm\ (\text{lumi.}) \pm \extrAfullWm\ (\text{extr.}) \, \text{pb}$,
\end{center}
and the total inclusive $Z$-boson cross-section in the combined electron and muon channels is:
\begin{center}
$\sigma^{\text{tot}}_{Z \to \ell \ell} = \valAfullZ  \pm \statAfullZ\
(\text{stat.}) \pm \sysAfullZ\ (\text{syst.}) \pm \lumiAfullZ\ (\text{lumi.})  \pm \extrAfullZ (\text{extr.}) \, \text{pb}.$
\end{center}
 
The results obtained, and the ratios and charge asymmetries constructed from them, are in agreement with theoretical calculations  based on NNLO QCD.

% End of text imported from the .//tex/Conclusion.tex input file

\clearpage
\appendix
\part*{Appendix}
\addcontentsline{toc}{part}{Appendix}
% The next lines are included from the .//tex/AppTheory.tex input file
\section{Theoretical predictions}
\label{app:theory}
This appendix presents the theoretical predictions used for comparison with the measurements in the main body of the paper. Table~\ref{tab:app:pred} shows
the predictions using the MMHT14nnlo68cl,  NNPDF31\_nnlo\_as\_0118,
ATLASepWZ12, HERAPDF2.0, and ABMP16 PDF sets with associated PDF
uncertainties.

\begin{table}[!h]
\caption{The predictions at NNLO in QCD, using the MMHT14nnlo68cl,
NNPDF31\_nnlo\_as\_0118, ATLASepWZ12, HERAPDF2.0, and ABMP16  PDF sets, for the  cross-sections measured in this study. \label{tab:app:pred} }
\begin{center}
\begin{tabular}{  c | c | c | c | c | c  }
& \multicolumn{5}{c}{Predicted cross-section $\pm$ PDF
uncertainty
[pb]} \\
 
\raisebox{6pt}{Quantity}           		 & MMHT14 &
NNPDF31
&
ATLASepWZ12
&
HERAPDF20
&
ABMP16 \\
\hline
\rule[-2mm]{0mm}{6mm}
$\sigma_{W^+}^{\text{fid}}$&$1397^{+29}_{-30}  $ &
$1428^{+24}_{-24}
$
& $1375^{+34}_{-30}  $ & $1429^{+91}_{-49}  $ &$1397^{+14}_{-14}  $\\
\rule[-2mm]{0mm}{6mm}
$\sigma_{W^-}^{\text{fid}}$&$773 ^{+17}_{-20} $ &
$778
^{+14}_{-14}
$
&$784
^{+19}_{-19}
$
&$806
^{+31}_{-21}
$ &$746 ^{+9}_{-9} $\\
\rule[-3mm]{0mm}{7mm}
$\sigma_{Z}^{\text{fid}}$&$199^{+4}_{-4}$ &
$203^{+4}_{-4}$
& $199^{+4}_{-4}$ &$199^{+11}_{-5}$ &$198.6^{+2.0}_{-2.0}$\\
\hline
\rule[-2mm]{0mm}{6mm}
$\sigma_{W^+}^{\text{tot}}$&$2138^{+43}_{-45} $ &
$2271^{+36}_{-36}
$
&$2086^{+54}_{-47}
$ &$2140^{+140}_{-70} $&$2214^{+21}_{-21} $\\
\rule[-2mm]{0mm}{6mm}
$\sigma_{W^-}^{\text{tot}}$&$1295^{+28}_{-33}$ &
$1330^{+22}_{-22}$ &$1296^{+48}_{-29}$ &$1338^{+52}_{-32}$&$1283^{+16}_{-16}$\\
\rule[-3mm]{0mm}{6mm}
$\sigma_{Z}^{\text{tot}}$&$308^{+6}_{-6}$
&$313^{+5}_{-5}$
&$308^{+6}_{-5}$ &$312^{+16}_{-7}$ &$305.7^{+3.0}_{-3.0}$\\
\end{tabular}
\end{center}
\end{table}

\clearpage
% End of text imported from the .//tex/AppTheory.tex input file
 
\section*{Acknowledgements}
% The next lines are included from the .//tex/Acknowledgements.tex input file
 
We thank CERN for the very successful operation of the LHC, as well as the
support staff from our institutions without whom ATLAS could not be
operated efficiently.
 
We acknowledge the support of ANPCyT, Argentina; YerPhI, Armenia; ARC, Australia; BMWFW and FWF, Austria; ANAS, Azerbaijan; SSTC, Belarus; CNPq and FAPESP, Brazil; NSERC, NRC and CFI, Canada; CERN; CONICYT, Chile; CAS, MOST and NSFC, China; COLCIENCIAS, Colombia; MSMT CR, MPO CR and VSC CR, Czech Republic; DNRF and DNSRC, Denmark; IN2P3-CNRS, CEA-DRF/IRFU, France; SRNSFG, Georgia; BMBF, HGF, and MPG, Germany; GSRT, Greece; RGC, Hong Kong SAR, China; ISF and Benoziyo Center, Israel; INFN, Italy; MEXT and JSPS, Japan; CNRST, Morocco; NWO, Netherlands; RCN, Norway; MNiSW and NCN, Poland; FCT, Portugal; MNE/IFA, Romania; MES of Russia and NRC KI, Russian Federation; JINR; MESTD, Serbia; MSSR, Slovakia; ARRS and MIZ\v{S}, Slovenia; DST/NRF, South Africa; MINECO, Spain; SRC and Wallenberg Foundation, Sweden; SERI, SNSF and Cantons of Bern and Geneva, Switzerland; MOST, Taiwan; TAEK, Turkey; STFC, United Kingdom; DOE and NSF, United States of America. In addition, individual groups and members have received support from BCKDF, CANARIE, CRC and Compute Canada, Canada; COST, ERC, ERDF, Horizon 2020, and Marie Sk{\l}odowska-Curie Actions, European Union; Investissements d' Avenir Labex and Idex, ANR, France; DFG and AvH Foundation, Germany; Herakleitos, Thales and Aristeia programmes co-financed by EU-ESF and the Greek NSRF, Greece; BSF-NSF and GIF, Israel; CERCA Programme Generalitat de Catalunya, Spain; The Royal Society and Leverhulme Trust, United Kingdom.
 
The crucial computing support from all WLCG partners is acknowledged gratefully, in particular from CERN, the ATLAS Tier-1 facilities at TRIUMF (Canada), NDGF (Denmark, Norway, Sweden), CC-IN2P3 (France), KIT/GridKA (Germany), INFN-CNAF (Italy), NL-T1 (Netherlands), PIC (Spain), ASGC (Taiwan), RAL (UK) and BNL (USA), the Tier-2 facilities worldwide and large non-WLCG resource providers. Major contributors of computing resources are listed in Ref.~\cite{ATL-GEN-PUB-2016-002}.
 
% End of text imported from the .//tex/Acknowledgements.tex input file

\printbibliography

\clearpage \input{atlas_authlist}

\end{document}

%% file: atlas_authlist.tex
% ATLAS Collaboration author list
% Reference date of STDM-2018-06 is 2019-02-12
% Author list last updated on date 27-FEB-20
% Data extracted on 27-Feb-2020 for paper reference STDM-2018-06
% at 1:40pm
 
\begin{flushleft}
{\Large The ATLAS Collaboration}

\bigskip

G.~Aad$^\textrm{\scriptsize 102}$,    
B.~Abbott$^\textrm{\scriptsize 129}$,    
D.C.~Abbott$^\textrm{\scriptsize 103}$,    
O.~Abdinov$^\textrm{\scriptsize 13,*}$,    
A.~Abed~Abud$^\textrm{\scriptsize 71a,71b}$,    
K.~Abeling$^\textrm{\scriptsize 53}$,    
D.K.~Abhayasinghe$^\textrm{\scriptsize 94}$,    
S.H.~Abidi$^\textrm{\scriptsize 167}$,    
O.S.~AbouZeid$^\textrm{\scriptsize 40}$,    
N.L.~Abraham$^\textrm{\scriptsize 156}$
% ...
\bigskip
\\

% ...
$^{13}$Institute of Physics, Azerbaijan Academy of Sciences, Baku; Azerbaijan.\\
% ...
$^{40}$Niels Bohr Institute, University of Copenhagen, Copenhagen; Denmark.\\
% ...
$^{53}$II. Physikalisches Institut, Georg-August-Universit\"{a}t G\"ottingen, G\"ottingen; Germany.\\
% ...
$^{71}$$^{(a)}$INFN Sezione di Pavia;$^{(b)}$Dipartimento di Fisica, Universit\`a di Pavia, Pavia; Italy.\\
% ...
$^{94}$Department of Physics, Royal Holloway University of London, Egham; United Kingdom.\\
% ...
$^{102}$CPPM, Aix-Marseille Universit\'e, CNRS/IN2P3, Marseille; France.\\
$^{103}$Department of Physics, University of Massachusetts, Amherst MA; United States of America.\\
% ...
$^{129}$Homer L. Dodge Department of Physics and Astronomy, University of Oklahoma, Norman OK; United States of America.\\
% ...
$^{156}$Department of Physics and Astronomy, University of Sussex, Brighton; United Kingdom.\\
% ...
$^{167}$Department of Physics, University of Toronto, Toronto ON; Canada.\\
% ...
$^{a}$ Also at Borough of Manhattan Community College, City University of New York, New York NY; United States of America.\\
$^{b}$ Also at Centre for High Performance Computing, CSIR Campus, Rosebank, Cape Town; South Africa.\\
% ...
$^{*}$ Deceased

\end{flushleft}

% Created with Glance <Atlas.Glance@cern.ch>